\documentclass[equations,11pt]{article}
\usepackage{amsfonts}
\usepackage{a4,tikz}
\usepackage{geometry}

\makeatletter
\newcommand\ackname{Acknowledgements}
\if@titlepage
  \newenvironment{acknowledgements}{%
      \titlepage
      \null\vfil
      \@beginparpenalty\@lowpenalty
      \begin{center}%
        \bfseries \ackname
        \@endparpenalty\@M
      \end{center}}%
     {\par\vfil\null\endtitlepage}
\else
  \newenvironment{acknowledgements}{%
      \if@twocolumn
        \section*{\abstractname}%
      \else
        \small
        \begin{center}%
          {\bfseries \ackname\vspace{-.5em}\vspace{\z@}}%
        \end{center}%
        \quotation
      \fi}
      {\if@twocolumn\else\endquotation\fi}
\fi
\makeatother

\usepackage{amsmath}
\usepackage{amssymb}
\usepackage{amsthm}
\usepackage{eucal}
 \usepackage[usenames,dvipsnames]{pstricks}
 \usepackage{epsfig}
 \usepackage{pst-grad} % For gradients
 \usepackage{pst-plot} % For axes
\renewcommand{\theequation}{\arabic{equation}}
\usepackage{amssymb,amsmath}
\usepackage{color}
\usepackage{accents}
\usepackage{texdraw}
\usepackage{eucal}
\usepackage{epic,epsfig}
\usepackage{graphicx}
\newtheorem{theorem}{Theorem}[section]
\newtheorem{lemma}{Lemma}[theorem]

\theoremstyle{definition}

\numberwithin{equation}{section}
\DeclareMathAccent{\wtilde}{\mathord}{largesymbols}{"65}
\DeclareMathAccent{\what}{\mathord}{largesymbols}{"62}

\def\m@th{\mathsurround=0pt}
\mathchardef\bracell="0365
\def\upbrall{$\m@th\bracell$}
\def\undertilde#1{\mathop{\vtop{\ialign{##\crcr
    $\hfil\displaystyle{#1}\hfil$\crcr
     \noalign
     {\kern1.5pt\nointerlineskip}
     \upbrall\crcr\noalign{\kern1pt
   }}}}\limits}
\def\m@th{\mathsurround=0pt}
\mathchardef\bracell="0365
\def\upbrall{$\m@th\bracell$}
\def\underhat#1{\mathop{\vtop{\ialign{##\crcr
    $\hfil\displaystyle{#1}\hfil$\crcr
     \noalign
     {\kern1.5pt\nointerlineskip}
     \upbrall\crcr\noalign{\kern1pt
   }}}}\limits}
\usepackage{pgf}
\usepackage{tikz}
\usepackage{verbatim}
\usepackage[utf8]{inputenc}
\usetikzlibrary{arrows,automata}
\def\theequation{\arabic{section}.\arabic{equation}}

\newcommand{\wh}{\widehat}
\newcommand{\wt}{\widetilde}

\def\hypotilde#1#2{\vrule depth #1 pt width 0pt{\smash{{\mathop{#2}
\limits_{\displaystyle\widetilde{}}}}}}
\def\hypohat#1#2{\vrule depth #1 pt width 0pt{\smash{{\mathop{#2}
\limits_{\displaystyle\widehat{}}}}}}

\newcommand{\bun}{\boldsymbol{1}}

\newcommand{\tbu}{\,^{t\!}{\bu}}
\newcommand{\bblu}{\begin{color}{blue}}
\newcommand{\bred}{\begin{color}{red}}
\newcommand{\ecl}{\end{color}}

\newcommand{\bB}{\boldsymbol{B}}
\newcommand{\bC}{\boldsymbol{C}}

\newcommand{\bK}{\boldsymbol{K}}

\newcommand{\bM}{\boldsymbol{M}}

\newcommand{\bS}{\boldsymbol{S}}

\newcommand{\aR}{\alpha}
\newcommand{\bb}{\beta}

\newcommand{\ven}{\varepsilon}

\newcommand{\dd}{\delta}
\newcommand{\sg}{\sigma}

\newcommand{\kp}{\kappa}
\newcommand{\ld}{\lambda}

\newcommand{\be}{\begin{equation}}
\newcommand{\ee}{\end{equation}}
\newcommand{\bea}{\begin{eqnarray}}
\newcommand{\eea}{\end{eqnarray}}
\newcommand{\bse}{\begin{subequations}}
\newcommand{\ese}{\end{subequations}}
\newcommand{\nn}{\nonumber}

\newcommand{\ol}{\overline}

\newcommand{\bu}{\boldsymbol{u}}

\newcommand{\brr}{\boldsymbol{r}}

\newcommand{\bs}{{\boldsymbol s}}

\begin{document}

\def\theequation{\arabic{section}.\arabic{equation}}

\newtheorem{thm}{Theorem}[section]
\newtheorem{lem}{Lemma}[section]
\newtheorem{defn}{Definition}[section]
\newtheorem{ex}{Example}[section]
\newtheorem{rem}{Remark}
\newtheorem{criteria}{Criteria}[section]
\newcommand{\ra}{\rangle}
\newcommand{\la}{\langle}
%\newcommand{\be}{\begin{equation}}
%\newcommand{\ee}{\end{equation}}
%\newcommand{\bea}{\begin{eqnarray}}
%\newcommand{\eea}{\end{eqnarray}}
%\newcommand{\bse}{\begin{subequations}}
%\newcommand{\ese}{\end{subequations}}
%\newcommand{\nn}{\nonumber}
%\input{tcilatex}

%%%%%%%%%%%%%%%%%%%%%%%%%%%%%%%%%%%%%%%%%%%%%%%%%%%%%%%%%%%%%%%%%%%%%%%%%%%%
\title{\textbf{Elliptic ($N,N^\prime$)-Soliton Solutions of the lattice KP Equation}}
\author{\\\\Sikarin Yoo-Kong$^{\dagger,\star,1 } $, Frank Nijhoff$^{\ddagger,2} $ \\
\small $^\dagger $\emph{Department of Physics, King Mongkut's University of Technology Thonburi,}\\
\small \emph{Thailand, 10140.}\\
\small $^\star $\emph{King Mongkut's University of Technology Thonburi Ratchaburi Campus, Thailand, 70150.}\\
\small $^\ddagger $\emph{School of Mathematics, Department of Applied Mathematics, University of Leeds,}\\
\small \emph{United Kingdom, LS2 9JT.}\\
\small  $^1$syookong@gmail.com,
\small $^2$nijhoff@maths.leeds.ac.uk}
\maketitle
%%%%%%%%%%%%%%%%%%%%%%%%%%%%%%%%%%%%%%%%%%%%%%%%%%%%%%%

%%%%%%%%%%%%%%%%%%%%%%%%%%%%%%%%%%%%%%%%%%%%%%%%%%%%%%%

\abstract
   Elliptic soliton solutions, i.e., a hierarchy of functions based on an elliptic seed solution, are constructed 
   using an elliptic Cauchy kernel, for integrable lattice equations of Kadomtsev-Petviashvili (KP) type. This 
   comprises the lattice KP, modified KP (mKP) and Schwarzian KP (SKP) equations as well as Hirota's bilinear KP 
   equation, and their successive continuum limits. The reduction to the elliptic soliton solutions of KdV type 
   lattice equations is also discussed.

%%%%%%%%%%%%%%%%%%%%%%%%%%%%%%%%%%%%%%%%%%%%%%%%%%%%%%%%

\section{The lattice KP and Hirota equations}
The study of the discrete versions of soliton systems, i.e., systems given by integrable partial difference equations, have become in recent years a focus of attention in the theory of integrable systems. Among those systems, the discrete analogue of Kadomtsev-Petviashvili (KP) equations which define in three dimensional lattice, seem to form a universal class of systems. The first equation of this type was found by Hirota in \cite{Hirota} and 
was referred to as DAGTE (Discrete analogue of generalised Toda equation) which is the bilinear equation
\be\label{eq:hirota}
\left(a_1e^{D_1}+a_2e^{D_2}+a_3e^{D_3} \right)\tau \cdot \tau =0\;,
\ee
where the Hirota operators $D_i$ produce finite forward-and backward shifts, when acting on a pair of functions, 
in the corresponding lattice direction, i.e.,  
\[ e^{D_1}f\cdot g=f(n_1+1,....)g(n_1-1,....)\  . \] 
Special reductions of this equation, are obtained  
when the coefficients $a_1$, $a_2$, $a_3$ satisfy the condition $a_1+a_2+a_3=0$, but the full equation 
is integrable in the sense of multidimensional consistency for arbitrary parameters, see \cite{ABS2}. 
Miwa \cite{Miwa} reparametrized the equation in that restricted case, and hence in that form it is often 
referred to as \textit{Hirota-Miwa equation}\footnote{Sometimes the full equation (representing the blinear 
discrete KP equation) is also (in our view erroneously) referred to as the Hirota-Miwa equation. In fact, 
in \cite{Miwa} only the restricted case was considered, and generalized to a four-term equation which is 
nowadays referred to as the Miwa equation.}. In this paper we will investigate a class of solutions of this and 
related three-dimensional lattice equations, comprising the following equations: 
\\
\\
\emph{The bilinear lattice KP equation}
\be\label{disbi} 
\ol{f}\wh{\wt{f}}+\wh{f}\wt{\ol{f}}-\wt{f}\wh{\ol{f}}=0\;.
\ee
\\
\\
\emph{The lattice KP equation}
\begin{equation}\label{dLKP}
(\wh{\ol{w}}-\wt{\ol{w}})(\ol{w}-\wt{w})=(\wh{\ol{w}}-\wh{\wt{w}})(\wh{w}-\wt{w})\;.
\end{equation}
\\
\\
\emph{The lattice modified KP equation}
\begin{eqnarray}\label{dMKP1}
\frac{\wh{ v}
-\wt{ v}}{\wh{\wt v}}
+\frac{\ol{ v}-\wh{ v}}{\wh{\ol{ v}}}
+\frac{\wt{v}-\ol{ v}}{\wt{\ol v}}=0\;.
\end{eqnarray}
\\
\\
\emph{The asymmetric modified KP}
\begin{eqnarray}\label{daKP}
\frac{\wh{\mathcal V}_0
-\wt{\mathcal V}_0}{\wh{\wt{\mathcal V}}_0}
+\frac{\ol{\mathcal V}_0}{\wh{\ol{\mathcal V}}_0}
-\frac{\ol{\mathcal V}_0}{\wt{\ol{\mathcal V}}_0}=0\;.
\end{eqnarray}
\\
\\ 
\emph{The lattice Schwarzian KP}
\begin{eqnarray}\label{dSKP1}
\frac{(\ol{z}-\wt{\ol{z}})(\wh{z}-\wh{\ol{z}})(\wt{z}-\wh{\wt{z}})}
{(\ol{z}-\wh{\ol{z}})(\wt{z}-\wt{\ol{z}})(\wh{z}-\wh{\wt{z}})}=1\;.
\end{eqnarray}
The notation in \eqref{disbi}-\eqref{dSKP1} is as follows: all dependent variables are functions defined on the multidimensional 
lattice with discrete coordinates $(n,m,h)\in\mathbb{Z}^3$~, e.g., $f=f(n,m,h)$, and the elementary shifts in the three 
discrete directions are denoted by $\wt{f}=f(n+1,m,h)$, $\wh{f}=f(n,m+1,h)$ and $\ol{f}=f(n,m,h+1)$. The first equation 
\eqref{disbi} is Hirota's DAGTE after a change of independent variables and a point transformation. The other 
KP type lattice equations \eqref{dLKP}-\eqref{dSKP1} where established in \cite{NCWQ} in a form containing parameters but which 
are equivalent to the normalised forms given above by point transformations. In particular, the Schwarzian KP \eqref{dSKP1} was first 
given in this normalised form in \cite{DN}, and was subsequently recovered in \cite{BK}, whilst its geometric significance was explored in \cite{KS2,KS,Doliwa1}. All equations \eqref{dLKP}--\eqref{dSKP1} appear as distinct parameter choices of a generalized lattice 
KP equation given in \cite{NCWQ}, whilst the \eqref{disbi} can be viewed as a potential equation for the asymmetric lattice MKP equation.   
With regard to terminology, we prefer to reserve the name lattice (potential) KP equation for \eqref{dLKP} rather than for \eqref{disbi} 
or for \eqref{eq:hirota} (as is commonly done in the literature), because it is actually the former equation that is more directly related 
to the actual continuous KP equation whilst the latter two equations are more related to the bilinear form of the KP equation. Similarly, 
the equations \eqref{dMKP1} and \eqref{dSKP1} can be shown to be related directly to the continuous (potential) modified and the 
Schwarzian KP respectively equations through continuum limits, cf. \cite{NCWQ,NCW,WC}, hence their names. 
It was recently established in \cite{ABS2} that this list of five canonical equations exhausts all possible cases of octahedral type lattice 
equations, up to equivalence, which are multidimensionally consistent when embedded in the four dimensional lattice. 

Some of these equations have been quite widely studied. In particular, eq. \eqref{eq:hirota} has arisen in many contexts over the last few years, 
notably in connection with affine Weyl group description of discrete Painlev\'e equations, \cite{NY}, and in connection with the representation 
theory of Kac-Moody algebras \cite{Kacbook}. An interesting connection was found with the spectrum of Bethe Ansatz states of quantum solvable 
models, the eigenvalues of which were shown to obey specific versions of Hirota's bilinear equation, cf. \cite{Lipan,Zabrodin1}. 
Furthermore, related to this is a connection found in \cite{Frank1} between, on the one hand, pole-type (in the sense of Airault, McKean and Moser 
\cite{Moser}) and soliton solutions of the lattice KP equation and a class of discrete-time many-body systems which effectively are gives rise 
to integrable correspondences (i.e., multi-valued dynamical maps) which are time-discretizations of the Ruijsenaars model \cite{R1,R2}. On the 
other hand the equations describing these same classical discrete dynamical systems coincide exactly with the Bethe-Ansatz equations 
for the excitations of quantum integrable models solvable by the quantum inverse scattering transform. It seems from all these intriguing 
connections that the lattice KP equations exhibit a  certain universality in their connection with various types of integrable models involving  
partial difference equations, classical many-body systems and quantum solvable systems, a full picture of which is not yet understood  
at this juncture.
 
In this paper, we consider a novel class of solution of these lattice equations namely soliton type solutions based on an elliptic seed 
solution. The construction exploits elliptic $N \times N^\prime$ Cauchy-matrices, and involve also $N^\prime \times N$ coefficient matrices, 
which in principle allows for a classification of the solutions according to the Schubert decomposition of a corresponding Grassmannian.  
Thus, our main result, which provides a general expression for what we accordingly call the elliptic $(N,N^\prime)$-soliton solutions, 
could form a basis for a similar analysis as was performed in the case of the continuous KP equation by Kodama and Chakravarty, \cite{Kodama}, 
to describe the soliton taxonomy of the KP equation. %Although, we do not intend to perform that full analysis 
%in the present paper, here we establish the general determinantal formulae on which such an analysis could be based in the discrete case. 
Furthermore, our approach yields the various Miura type relations between the different equations and allows to study the various intermediate 
continuum limits, as well as reductions to lower dimensional equations, in a systematic way.

%

%
%
%
%
%%%%%%%%%%%%%%%%%%%%%%%%%%%%%%%%%%%%%%%%%%%%%%%%%%%%%%%%%%%%%%%%%%%%%%%%%
\section{Cauchy matrix scheme}
We will develop now a scheme along the lines of the paper \cite{NAH,FrankJames} where Atkinson and one of the present authors developed the elliptic soliton 
solutions for the ABS (Adler/Bobenko/Suris, cf. \cite{ABS}) list of two-dimensional lattice equations, based on elliptic Cauchy matrices. 
That construction provided solutions for all equations of the ABS list up to $Q_3$, whereas the case of $Q_4$ was treated separately using a different 
approach, cf. \cite{Q4}.  

In this Section we derive the basic relations, and in the next Section we 
will use these relations to find a general elliptic $(N,N^\prime)$-soliton solution for KP.

\subsection{Basic ingredients}

At this point we find it useful to introduce the Lam\'e function 
\be\label{eq:Psi}
\Psi_\xi(\kp)=: \Phi_\xi(\kp)\,e^{-\zeta(\xi)\kp}\   ,  
\ee
where
\be\label{eq:Phia}
\Phi_\xi(\kp)= \frac{\sigma(\kp+\xi)}{\sigma(\xi)\sigma(\kp)} ,  
\ee
with $\sigma$ is the sigma function and we have included an exponential factor together with the $\Phi$-function, breaking the symmetry between 
the argument of the function and the suffix. Although most of the results of this paper can be obtained in terms of the 
$\Phi$-function alone\footnote{The inclusion of the exponential factor amounts to a specific gauge transformation on the 
quantities defined later on in the construction, and hence could be removed without affecting the main results.}, i.e. without adopting the exponential factor, 
the exponential factor does provide a certain regularisation of the functions involved which may affect their 
analytic behaviour. The basic identities for the $\Psi$ function are the following:
\bse\label{eq:Psiids}\bea 
  \Psi_\xi(\kp)\Psi_\dd (\ld)&=&e^{\eta_\dd\kp} \Psi_{\xi+\dd}(\kp)\Psi_\dd(\ld-\kp)+ 
e^{\eta_\dd\ld} \Psi_\xi(\kp-\ld)\Psi_{\xi+\dd}(\ld)\   , \label{eq:Psiidsa} \\ 
 \Psi_\xi(\kp)\Psi_\dd(\kp) &=& e^{\eta_\dd\kp}\Psi_{\xi+\dd}(\kp)\,\left[\zeta(\xi)+\zeta(\dd)+\zeta(\kp)-\zeta(\xi+\dd+\kp) \right]\   , \label{eq:Psiidsb}\\ 
\Psi_\xi(\kp)\Psi_\xi(\ld) &=& \Psi_{\xi}(\kp+\ld)\,\left[\zeta(\xi)+\zeta(\kp)+\zeta(\ld)-\zeta(\xi+\kp+\ld) \right]\   , 
\label{eq:Psiidsc}
\eea\ese 
in which we have introduced 
\be\label{eq:eta}
\eta_\dd=\eta_\dd(\xi)=\zeta(\xi+\dd)-\zeta(\xi)-\zeta(\dd)=\frac{1}{2}\,\frac{\wp'(\xi)-\wp'(\dd)}{\wp(\xi)-\wp(\dd)}\  . 
\ee 
Furthermore, we have the symmetry: ~$\Psi_\dd(-\kp)=-\Psi_{-\dd}(\kp)$~. 

The starting point for our construction is the ``bare" non-autonomous Cauchy matrix~ 
\be\label{eq:Cauchy} 
\boldsymbol{M^0}=\left(M^0_{i,j}\right)_{i,j=1,\dots,N}\quad ,\quad  M^0_{i,j}(\xi)\equiv \Psi_\xi(\kp_i+\kp_j^\prime)\  , 
\ee 
depending on a variable $\xi$ depending linearly on independent variables $n$, $m$, namely ~$\xi\equiv\xi_0+n\dd+m\ven+h\lambda$~~, with 
$\dd$, $\ven$ and $\lambda$ being the corresponding \textit{lattice parameters}. We will assume that the set rapidity parameters $\{\kp_i, i=1,\dots,N\}$
is such that $\kp_i+\kp_j^\prime\neq 0$ (modulo the period lattice of the $\sg$-function).

Furthermore, defining
\be\label{eq:pq} p_\kp=\Psi_\dd(\kp)\quad,\quad q_\kp=\Psi_\ven(\kp),\quad l_\kp=\Psi_\lambda(\kp),  \ee
and setting $\kp=\pm\kp_i$, $\kp^\prime=\pm\kp_j^\prime$, we can derive   
from the basic addition formula \eqref{eq:Psiidsa} the following dynamical properties of the elliptic Cauchy matrix:
\begin{eqnarray*}
M_{i,j}^0p_{\kp_j^\prime} &=& \Psi_\xi(\kp_i+\kp_j^\prime)\Psi_\dd(\kp_j^\prime) \\ 
&=& e^{\eta_\dd(\kp_i+\kp_j^\prime)}\Psi_{\xi+\dd}(\kp_i+\kp_j^\prime)\Psi_\dd(-\kp_i) + 
e^{\eta_\dd\kp_j^\prime}\Psi_{\xi+\dd}(\kp_j^\prime)\Psi_\xi(\kp_i) \\ 
&=& \wt{M}_{i,j}^0p_{-\kp_i}e^{\eta_\dd(\kp_i+\kp_j^\prime)}+e^{\eta_\dd\kp_j^\prime}\Psi_{\xi+\dd}(\kp_j^\prime)\Psi_\xi(\kp_i)\  . 
\end{eqnarray*}
%
%Similarly, we have 
%$$ p_{-\kp_i}\wt{M}_{i,j}^0=M_{i,j}^0p_{\kp_j^\prime}e^{-\eta_\dd(\kp_i+\kp_j^\prime)}-e^{-\eta_\dd\kp_i}\Psi_\xi(\kp_i)\Psi_{\xi+\dd}(\kp_j^\prime)\  .  $$ 
%
We introduce now the plane-wave factors (i.e., discrete exponential functions)
\bea
 \rho(\kp)&=&\left(e^{-\zeta(\dd)\kp}\,p_{-\kp}\right)^{~n~} 
\left(e^{-\zeta(\ven)\kp}\,q_{-\kp}\right)^{~m~}\left(e^{-\zeta(\lambda)\kp}\,l_{-\kp}\right)^{~h~}e^{\zeta(\xi)\kp}\rho_{0,0,0}(\kp), \label{eq:rho1}\\
 \nu(\kp^\prime)&=&\left(e^{\zeta(\dd)\kp^\prime}\,p_{\kp^\prime}\right)^{~-n~} 
\left(e^{\zeta(\ven)\kp^\prime}\,q_{\kp^\prime}\right)^{~-m~}\left(e^{\zeta(\lambda)\kp^\prime}\,l_{\kp^\prime}\right)^{~-h~}e^{\zeta(\xi)\kp^\prime}\rho_{0,0,0}(\kp^\prime),\label{eq:rho22}
\eea 
and for the specific values $\kp=\kp_i$, ~$\rho_i:=\rho(\kp_i)$~  and $\kp^\prime=\kp_j^\prime$, ~$\nu_j:=\rho(\kp_j^\prime)$~  obeying the shift relations
\be\label{eq:rhoshifts}
\frac{\wt{\rho}_i}{\rho_i} = e^{\eta_\dd\kp_i}p_{-\kp_i}\quad, \quad 
\frac{\wh{\rho}_i}{\rho_i} = e^{\eta_\ven\kp_i}q_{-\kp_i}\quad, \quad 
\frac{\ol{\rho}_i}{\rho_i} = e^{\eta_\lambda\kp_i}l_{-\kp_i}\\  , 
\ee 
\be\label{eq:nushifts}
\frac{\wt{\nu}_j}{\nu_j} = e^{\eta_\dd\kp_j^\prime}(p_{\kp_j^\prime})^{-1}\quad, \quad 
\frac{\wh{\nu}_j}{\nu_j} = e^{\eta_\ven\kp_j^\prime}(q_{\kp_j^\prime})^{-1}\quad, \quad 
\frac{\ol{\nu}_j}{\nu_j} = e^{\eta_\lambda\kp_j^\prime}(l_{\kp_j^\prime})^{-1}\\  , 
\ee 
where the superscripts ``$\;\wt{\phantom{a}}\;$", ``$\;\wh{\phantom{a}}\;$" and ``$\;\ol{\phantom{a}}\;$" denote the lattice shifts related to the shift by one unit in the variables 
$n$, $m$ and $h$ respectively, making use also of the relations ~$\wt{\xi}=\xi+\dd$~,~$\wh{\xi}=\xi+\ven$~ and ~$\ol{\xi}=\xi+\lambda$~ which sit inside the coefficients 
$\eta_\dd$, $\eta_\ven$ and $\eta_\lambda$.
\\
\\
Now we can introduce the $N$- and $N^\prime$-component vectors 
\be\label{eq:rs} 
\brr=\left(\rho_i\Psi_\xi(\kp_i)\right)_{i=1,\dots,N}\quad,\quad \bs=\left(\nu_j\Psi_\xi(\kp_j^\prime)\right)_{j=1,\dots,N^\prime} 
\ee
where ~$\rho_i=\rho_{n,m,h}(\kp_i)$~~ and ~$\nu_j=\nu_{n,m,h}(\kp_j^\prime)$~~  in terms of which 
we can define now the ``dressed" Cauchy matrix:
\be\label{eq:MCauchy} \bM=\left(\bM_{i,j}\right)_{i,j=1,\dots,N}\quad, \quad \bM_{i,j}=\rho_i\bM_{i,j}^0\nu_j\  .
\ee  
As a consequence of the relations given earlier, and employing the definitions of the plane-wave factors, 
we can now describe the discrete dynamics as follows:
\begin{lemma}
The dressed Cauchy matrix $\bM$, as defined in \eqref{eq:MCauchy}, obeys the following linear relations under elementary shifts 
of the independent variables $n$ 
\begin{eqnarray}\label{eq:qM}
 \wt{\bM}=\bM-\brr\,\wt{s}^T \label{eq:pMa}, 
\end{eqnarray} 
and under shifts of the variable $m$ the similar relations:
\begin{eqnarray}\label{eq:pM}
 \wh{\bM}=\bM-\brr\,\wh{s}^T \label{eq:pMb}, 
\end{eqnarray}
and under shifts of the variable $h$ the similar relations:
\begin{eqnarray}\label{eq:gM}
\ol{\bM}=\bM-\brr\,\ol{s}^T \label{eq:pMb}, 
\end{eqnarray} 
where, as before, $\wt{\bM}$, $\wh{\bM}$ and $\ol{\bM}$ denote the shifted Cauchy matrices. 
\end{lemma}
In what follows we will employ the relations \eqref{eq:qM}, \eqref{eq:pM} and \eqref{eq:gM} to obtain nonlinear shift relations for 
specific objects defined in terms of the Cauchy matrix $\bM$. 
%%%%%%%%%%%%%%%%%%%%%%%%%%%%%%%%%%%%%%%%%%%%%%%%%%%%%%%%%%%%%%%%%%%%%%%%%%%%%%%%%%%%%%%%%%%%%%%%%%%%%%
\subsection{The $\tau$-function and related basic objects} 

Introduce now the ~$\tau$-function~~:
\be\label{eq:tau} 
\tau=\tau_{n,m,h}=\det_{N\times N}\left(\boldsymbol{1}+\boldsymbol{MC}\right)=\det_{N^\prime\times N^\prime}\left(\boldsymbol{1}+\boldsymbol{CM}\right) \  , 
\ee  
where, since $\boldsymbol{M}$ is in general not a square matrix, the $N^\prime\times N$ constant matrix $\boldsymbol{C}$ is introduced to compensate for the discrepancy. The matrix $\boldsymbol{1}$ is either the $N\times N$ or $N^\prime\times N^\prime$ unit matrices respectively in both determinants. The latter identity is a consequence of the general Weinstein-Aronszajn formula:
\be\label{eq:WA} 
\det_{N\times N}\left(\boldsymbol{1}+\sum_{l=1}^{N^\prime}\boldsymbol{m}_l\boldsymbol{c}_l\right)=
\det_{N^\prime\times N^\prime}\left(\boldsymbol{1}+\boldsymbol{c}_l^t\boldsymbol{m}_k\right) \  , 
\ee  
where $\boldsymbol{m}_l=(M_{i,l})_{i=1,...,N}$ and $\boldsymbol{c}^t_l=(C_{i,l})_{j=1,...,N}$ are the $N^\prime$ column-resp. row vectors from the matrices $\boldsymbol M$ and 
$\boldsymbol C$.

From the definition of the $\tau$-function and the relations for the dressed Cauchy matrix $\bM$ we obtain:
\begin{eqnarray*}
\wt{\tau} &=& \det_{N\times N}\left( \bun+\wt{\bM}\bC\right)=
\det_{N\times N}\left\{ \bun+(\bM-\boldsymbol{r} \wt{\boldsymbol{s}}^T)\bC \right\} \\  
&=& \det_{N\times N}\left\{ (\bun + \boldsymbol{MC}) \left[ \bun - (\bun+\boldsymbol{MC})^{-1}{\brr} \wt{\bs}^T\bC \right]  \right\}\nn\\
&=&\tau\det_{N\times N}\left\{\left[ \bun - (\bun+\boldsymbol{MC})^{-1}{\brr} \wt{\bs}^T\bC \right]  \right\}.
\end{eqnarray*} 
Introducing 
\be\label{eq:Chi} 
\chi_{\aR,\bb}=\chi_{\aR,\bb}(\xi)\equiv \zeta(\aR)+\zeta(\bb)+\zeta(\xi)-\zeta(\xi+\aR+\bb)\ ,
\ee
and using the fact that
\begin{eqnarray} 
\wt{\boldsymbol{s}}=e^{\eta_\dd\bK^\prime}(p_{\bK^\prime})^{-1}\frac{\Psi_{\wt{\xi}}(\bK^\prime)}{\Psi_\xi(\bK^\prime)}\,
\boldsymbol{s}=\left[ \chi_{\delta,\bK^\prime}\right]^{-1}\boldsymbol{s},
\end{eqnarray}
where $p_{\bK^\prime}$ is the diagonal matrix with entries $p_{\kappa_j^\prime}$
and $\bK^\prime=\mbox{diag}(\kappa_1^\prime,...,\kappa_N^\prime)$.
Here and in what follows the notation $\chi_{\delta,\bK^\prime}$ denotes the diagonal matrix with entries 
$\chi_{\delta,\kappa_j^\prime}$, ($j=1,\dots,N^\prime$).

Now we have
\be\label{eq:ff}
\frac{\wt \tau}{\tau}=1-{\bs}^T\left[ \chi_{\delta,\bK^\prime}\right]^{-1}\bC\left[ 1+\bM\bC\right]^{-1}\brr=W_{\delta}.
\ee
The reverse fraction of Eq. \eqref{eq:ff} can be computed by processing the same computation
\begin{eqnarray*}
\tau &=& \det_{N\times N}\left( \bun+\bM\bC\right)=
\det_{N\times N}\left\{ \bun+(\wt{\bM}+\boldsymbol{r} \wt{\boldsymbol{s}}^T)\bC \right\} \\  
&=& \det_{N\times N}\left\{ (\bun + \boldsymbol{\wt{M}C}) \left[ \bun + (\bun+\boldsymbol{\wt{M}C})^{-1}{\brr} \wt{\bs}^T\bC \right]
  \right\}\nn\\
&=&\wt{\tau}\det_{N\times N}\left\{\left[ \bun + (\bun+\boldsymbol{\wt{M}C})^{-1}{\brr} \wt{\bs}^T\bC \right]
  \right\}.
\end{eqnarray*} 
Using also the fact that
\begin{eqnarray} 
\wt{\boldsymbol{r}}=e^{\eta_\dd\bK}p_{-\bK}\frac{\Psi_{\wt{\xi}}(\bK)}{\Psi_\xi(\bK)}\,
\boldsymbol{r}=-\left[ \wt{\chi}_{-\delta,\bK}\right]\boldsymbol{r},
\end{eqnarray}
where $\bK=\mbox{diag}(\kappa_1,...,\kappa_N)$ and then we have
\be\label{eq:ff2}
\frac{ \tau}{\wt \tau}=1-\wt{\bs}^T\left[ 1+\wt{\bM}\bC\right]^{-1}\bC\left[ \wt{\chi}_{-\delta,\bK}\right]^{-1}\wt{\brr}=\wt{V}_{-\delta}.
\ee
For arbitrary parameter $\alpha$ we introduce
\begin{eqnarray}
W_{\alpha} &=& 1-{\bs}^T\left[ \chi_{\alpha,\bK^\prime}\right]^{-1}\bC\left[ 1+\bM\bC\right]^{-1}\brr,\\
V_{\alpha} &=& 1-{\bs}^T\left[ 1+\bC\bM\right]^{-1}\bC\left[ \chi_{\alpha,\bK}\right]^{-1}\brr,
\end{eqnarray}
where we conclude that
\be\label{tau14}
\frac{\wt \tau}{\tau}=W_\delta=\frac{1}{\wt{V}_{-\delta}}.
\ee
We will now proceed the derivation of the dynamical relations of $W_\delta$ and $V_\delta$.
%%%%%%%%%%%%%%%%%%%%%%%%%%%%%%%%%%%%%%%%%%%%%%%%%%%%%%%%%%%%%%%%%%%%%%%%%%%%%%%%%%%%%%%%%%%%%%
\subsection{Basic linear relations}

In order to derive relations for the objects $V_\aR$ and $W_{\aR}$ we need to introduce the $N$- and $N^\prime-$component 
column- resp. row vectors: 
\bse\label{eq:butbu}\begin{eqnarray}
\bu_\aR &=& (\bun+\bM\bC)^{-1}(\chi_{\aR,\bK})^{-1}\brr \  , \label{eq:bu} \\ 
\tbu_\bb &=& \bs^T\,(\chi_{\bb,\bK^\prime})^{-1}(\bun+\bC\bM)^{-1} \  .   \label{eq:tbu}
\end{eqnarray}\ese 
Using these equations, we can write the functions $V$ and $W$ in the forms
\bse\label{eq:VWu}\begin{eqnarray}
V_{\aR}&=&1-\bs^{T}\bC\bu_{\aR}\;, \label{eq:bu} \\ 
W_{\bb}&=&1-\tbu_\bb\bC\boldsymbol{r}\; .   \label{eq:tbu}
\end{eqnarray}\ese 
Performing the following calculation:
\begin{eqnarray*}
&&\wt{\bu}_\aR= (\bun+\wt{\bM}\bC)^{-1}(\wt{\chi}_{\aR,\bK})^{-1}\wt{\brr}=(\bun+\wt{\bM}\bC)^{-1}(\wt{\chi}_{\aR,\bK})^{-1}
e^{\eta_\dd\bK}\frac{\Psi_{\wt{\xi}}(\bK)p_{-\bK}}{\Psi_\xi(\bK)}\brr \\ 
&& \Rightarrow\quad (\bun+\wt{\bM}\bC)\wt{\bu}_\aR= \frac{\Phi_{\wt{\xi}+\aR}(\bK)}
{\Phi_{\wt{\xi}}(\bK)\,\Phi_{\aR}(\bK)}\,
e^{\eta_\dd\bK}\frac{\Psi_\dd(-\bK)\,\Psi_{\wt{\xi}}(\bK)}{\Psi_\xi(\bK)}\,\brr  \nn\\
 && \Rightarrow\quad \left[(\bun+\bM\bC)-\brr\,\wt{\bs}^T\bC\right]\,\wt{\bu}_\aR = 
-\frac{\zeta(\bK)+\zeta(\wt{\xi}+\aR)-\zeta(\dd)-\zeta(\bK+\xi+\aR)}{\zeta(\bK)+\zeta(\xi)+\zeta(+\aR)-\zeta(\bK+\xi+\aR)}\,\brr \\ 
&& \qquad\qquad = \left(-1+\frac{\zeta(+\aR)+\zeta(\xi)-\zeta(\wt{\xi}+\aR)+\zeta(\dd)}
{\zeta(\bK)+\zeta(\xi)+\zeta(+\aR)-\zeta(\bK+\xi+\aR)}\right)\,\brr
\end{eqnarray*}
Multiplying both sides by ~$(\bun+\bM\bC)^{-1}$~ and introducing the vector
\be\label{eq:bu0} 
\bu_0\equiv(\bun+\bM\bC)^{-1}\brr\  , 
\ee 
we get the relation
\be\label{eq:burel} 
\wt{\bu}_\aR=-\bu_0\wt{V}_\aR+\chi_{\aR,\dd}\bu_\aR\  .
\ee
A similar set of relations can be derived for the adjoint vectors \eqref{eq:tbu} which involves the 
adjoint vector to \eqref{eq:bu0}, namely 
\be\label{eq:tbu0} 
\tbu_0\equiv\bs^T\,(\bun+\bC\bM)^{-1}\  , 
\ee 
and obviously these relations all have their counterparts involving the other lattice shift related to shifts in 
the discrete independent variables $m$ and $h$ instead of $n$.   

Summarising the results of these derivations, we have the following statement:
\begin{lemma}
The $N$- and $N^\prime$-component vectors given in \eqref{eq:butbu}, together with the ones given in \eqref{eq:bu0} and \eqref{eq:tbu0} obey 
the following set of linear shift relations 
\bse\begin{eqnarray}
 &&\wt{\bu}_\aR=-\bu_0\wt{V}_\aR+\chi_{\aR,\dd}\bu_\aR\  , \label{wtu} \\ 
 && \tbu_\bb = W_{\bb}\,^{t\!}\wt{\bu}_0 - \wt{\chi}_{\bb,-\dd}\,^{t\!}\wt{\bu}_\bb, \label{wtu2}
\end{eqnarray}\ese 
and a similar set of relations involving the shifts in the variable $m$ is given
\bse\begin{eqnarray}
 &&\wh{\bu}_\aR=-\bu_0\wh{V}_\aR+\chi_{\aR,\ven}\bu_\aR\  ,  \\ 
 && \tbu_\bb = W_{\bb}\,^{t\!}\wh{\bu}_0 - \wh{\chi}_{\bb,-\ven}\,^{t\!}\wh{\bu}_\bb, 
\end{eqnarray}\ese
and a similar set of relations involving the shifts in the variable $h$ is given
\bse\begin{eqnarray}
 &&\ol{\bu}_\aR=-\bu_0\ol{V}_\aR+\chi_{\aR,\lambda}\bu_\aR\  ,  \\ 
 && \tbu_\bb = W_{\bb}\,^{t\!}\ol{\bu}_0 - \ol{\chi}_{\bb,-\lambda}\,^{t\!}\ol{\bu}_\bb.\label{eq:tu} 
\end{eqnarray}\ese
\end{lemma}
%
%%%%%%%%%%%%%%%%%%%%%%%%%%%%%%%%%%%%%%%%%%%%%%%%%%%%%%%%%%%%%%%%
\subsection{Basic nonlinear relations}
The nonlinear relations can be obtained by multiply $\bs^T\bC$ on the left hand side of \eqref{wtu}. We have
\begin{eqnarray}
\bs^T\bC\wt{\bu}_\aR&=&-\bs^T\bC\bu_0\wt{V}_\aR+\chi_{\aR,\delta}\bs^T\bC\bu_{\aR}\nn\\
\wt{\bs}^T\chi_{\delta,\bK^\prime}\bC\wt{\bu}_\aR&=&-w_0\wt{V}_\aR+\chi_{\aR,\delta}(1-V_{\aR})\;,\label{Wtild}
\end{eqnarray}
where $w_0=\bs^T\bC\bu_0$. The same relations can be obtained in the same way for other shift directions
\begin{eqnarray}
\wh{\bs}^T\chi_{\varepsilon,\bK^\prime}\bC\wh{\bu}_\aR&=&-w_0\wh{V}_\aR+\chi_{\aR,\varepsilon}(1-V_{\aR})\;,\label{What}\\
\ol{\bs}^T\chi_{\lambda,\bK^\prime}\bC\ol{\bu}_\aR&=&-w_0\ol{V}_\aR+\chi_{\aR,\lambda}(1-V_{\aR})\;.\label{Wbar}
\end{eqnarray}
Combining \eqref{Wtild} and \eqref{What}, we obtain
\begin{eqnarray}\label{Vw}
\wh{\wt{V}}_{\aR}(\wh{w}-\wt{w})=\wh\chi_{\aR,\delta}\wh{V}_{\aR}-\wt\chi_{\aR,\varepsilon}\wt{V}_\aR\;,
\end{eqnarray}
where $w=\zeta(\xi)-w_0-n\zeta(\delta)-m\zeta(\varepsilon)-h\zeta(\gamma)$ .
\\
\\
Similarly, multiplying the right hand side of \eqref{wtu2} with $\bC\wt{\boldsymbol{r}}$, we obtain
\begin{eqnarray}\label{Ww}
{W}_{\bb}(\wt{w}-\wh{w})=\wt\chi_{\bb,-\delta}\wt{W}_{\bb}-\wh\chi_{\bb,-\varepsilon}\wh{W}_\bb\;.
\end{eqnarray}
We now can write $\wt{w}-\wh{w}$ in the following
\begin{eqnarray}\label{WVw}
(\wh{w}-\wt{w})&=&\frac{1}{\wh{\wt{V}}_{\aR}}\left(\frac{\Phi_\aR(\wh\xi)\Phi_\aR(\delta)}{\Phi_\aR(\wh\xi+\delta)}\wh{V}_{\aR}
-\frac{\Phi_\aR(\wt\xi)\Phi_\aR(\varepsilon)}{\Phi_\aR(\wt\xi+\varepsilon)}\wt{V}_\aR\right)\;,\\
&=&\frac{1}{W_{\bb}}\left(\frac{\Phi_\bb(\wh\xi)\Phi_\bb(-\varepsilon)}{\Phi_\bb(\wh\xi-\varepsilon)}\wh{W}_{\bb}
-\frac{\Phi_\bb(\wt\xi)\Phi_\bb(-\delta)}{\Phi_\bb(\wt\xi-\delta)}\wt{W}_\bb\right)\;.
\end{eqnarray}
Introducing new variables $\mathcal{V}_\aR=\Phi_{\aR}(\xi)V_\aR$ and
$\mathcal{W}_\bb=\Phi_{\bb}(\xi)W_\bb$, we now have
\begin{eqnarray}\label{WVw2}
(\wh{w}-\wt{w})&=&\frac{1}{\wh{\wt{\mathcal V}}_{\aR}}\left(p_\aR\wh{\mathcal V}_{\aR}
-q_\aR\wt{\mathcal V}_\aR\right)
=\frac{1}{\mathcal W_{\bb}}\left(p_{-\bb}\wh{\mathcal W}_{\bb}
-q_{-\bb}\wt{\mathcal W}_\bb\right)\;.
\end{eqnarray}
This equation has its counterpart involving the other lattice directions namely
\begin{subequations}
\begin{eqnarray}
(\wh{w}-\ol{w})&=&\frac{1}{\wh{\ol{\mathcal V}}_{\aR}}\left(l_\aR\wh{\mathcal V}_{\aR}
-q_\aR\ol{\mathcal V}_\aR\right)
=\frac{1}{\mathcal W_{\bb}}\left(l_{-\bb}\wh{\mathcal W}_{\bb}
-q_{-\bb}\ol{\mathcal W}_\bb\right)\;.\label{WVw3}\\
(\ol{w}-\wt{w})&=&\frac{1}{\ol{\wt{\mathcal V}}_{\aR}}\left(p_\aR\ol{\mathcal V}_{\aR}
-l_\aR\wt{\mathcal V}_\aR\right)
=\frac{1}{\mathcal W_{\bb}}\left(p_{-\bb}\ol{\mathcal W}_{\bb}
-l_{-\bb}\wt{\mathcal W}_\bb\right)\;.\label{WVw4}
\end{eqnarray}
\end{subequations}
Using \eqref{WVw2}, \eqref{WVw3} and \eqref{WVw4}, we have
\begin{eqnarray}\label{MKP1}
\frac{p_\aR\wh{\mathcal V}_{\aR}
-q_\aR\wt{\mathcal V}_\aR}{\wh{\wt{\mathcal V}}_{\aR}}
+\frac{q_\aR\ol{\mathcal V}_\aR-l_\aR\wh{\mathcal V}_{\aR}}{\wh{\ol{\mathcal V}}_{\aR}}
+\frac{l_\aR\wt{\mathcal V}_\aR-p_\aR\wh{\mathcal V}_{\aR}}{\wt{\ol{\mathcal V}}_{\aR}}=0\;,
\end{eqnarray}
or equivalently
\begin{eqnarray}\label{MKP2}
\frac{p_{-\bb}\wt{\ol{\mathcal W}}_{\bb}
-q_{-\bb}\wh{\ol{\mathcal W}}_\bb}{\ol{\mathcal W}_{\bb}}
+\frac{q_{-\bb}\wh{\wt{\mathcal W}}_\bb-l_{-\bb}\wt{\ol{\mathcal W}}_{\bb}}{\wt{\mathcal W}_{\bb}}
+\frac{l_{-\bb}\wh{\ol{\mathcal W}}_\bb-p_{-\bb}\wh{\wh{\mathcal W}}_{\bb}}{\wh{\mathcal W}_{\bb}}=0\;.
\end{eqnarray}
These two equations are the ``\emph{modified lattice KP}".
\\
\\
Taking $\aR=-\delta$, \eqref{MKP1} becomes
\begin{eqnarray}\label{aMKP1}
\frac{q_{-\delta}\ol{\mathcal V}_{-\delta}-l_{-\delta}\wh{\mathcal V}_{{-\delta}}}{\wh{\ol{\mathcal V}}_{{-\delta}}}
+\frac{l_{-\delta}\wt{\mathcal V}_{-\delta}}{\wt{\ol{\mathcal V}}_{{-\delta}}}-\frac{
q_{-\delta}\wt{\mathcal V}_{-\delta}}{\wh{\wt{\mathcal V}}_{{-\delta}}}=0\;.
\end{eqnarray}
Taking $\beta=\delta$, \eqref{MKP2} becomes
\begin{eqnarray}\label{aMKP2}
\frac{q_{-\delta }\wh{\wt{\mathcal W}}_\delta-l_{-\delta}\wt{\ol{\mathcal W}}_{\delta}}{\wt{\mathcal W}_{\delta}}
+\frac{l_{-\delta}\wh{\ol{\mathcal W}}_\delta}{\wh{\mathcal W}_{\delta}}-\frac{
q_{-\delta}\wh{\ol{\mathcal W}}_\delta}{\ol{\mathcal W}_{\delta}}=0\;.
\end{eqnarray}
\eqref{aMKP1} and \eqref{aMKP2} are the ``\emph{asymmetric modified KP}". 
%We see that these equations involve two of functions $q$ and $l$. If we take $\alpha=-\varepsilon $ and $\alpha=-\lambda $ we obtain equations involving functions $p,l$ and $p,q$ respectively. 
\\
\\
Furthermore, for $\aR=-\delta$, we find that
\begin{eqnarray}
(\wh{w}-\wt{w})&=&-q_{-\delta}\wt{\mathcal V}_{-\delta}/\wh{\wt{\mathcal V}}_{-\delta}\;,\label{WVw2d1}\\
(\ol{w}-\wt{w})&=&-l_{-\delta}\wt{\mathcal V}_{-\delta}/\wt{\ol{\mathcal V}}_{-\delta}\;.\label{WVw2d2}
\end{eqnarray}
The combination of these two equations yields
\begin{equation}\label{LKP4}
(\wh{\ol{w}}-\wt{\ol{w}})(\ol{w}-\wt{w})=(\wh{\ol{w}}-\wh{\wt{w}})(\wh{w}-\wt{w})\;,
\end{equation}
which is the ``\emph{lattice KP equation}".
\\
\\
Using \eqref{tau14}, we can write \eqref{WVw2d1} and \eqref{WVw2d2} in the forms
\begin{eqnarray}
(\wh{w}-\wt{w})&=&-q_{-\delta}\frac{\tau}{\wt{\tau}}\frac{\wh{\wt{\tau}}}{\wh{\tau}}\;,\label{wtau1}\\
(\ol{w}-\wt{w})&=&-l_{-\delta}\frac{\tau}{\wt{\tau}}\frac{\wt{\ol{\tau}}}{\ol{\tau}}\;.\label{wtau2}
\end{eqnarray}
From \eqref{WVw3}, if we take $\aR=-\varepsilon $ we also have
\begin{eqnarray}
(\wh{w}-\ol{w})&=&-q_{-\varepsilon }\frac{\tau}{\wh{\tau}}\frac{\wh{\ol{\tau}}}{\ol{\tau}}\;.\label{wtau3}
\end{eqnarray}
The combination of \eqref{wtau1}, \eqref{wtau2} and \eqref{wtau3} gives
\begin{eqnarray}\label{BKP}
l_{-\delta}\wh{\tau}\ol{\wt{\tau}}+q_{-\varepsilon }\wt\tau\wh{\ol{\tau}}-q_{-\delta}\ol\tau\wh{\wt{\tau}}=0\;,
\end{eqnarray}
which is actually the ``\emph{Hirota's DAGTE}". In the rational case, the summation of coefficients would add up to be zero. In the elliptic case, this condition is no longer to be the case.

%%%%%%%%%%%%%%%%%%%%%%%%%%%%%%%%%%%%%%%%%%%%%%%%%%%%%%%%%%%%%%%%%%%%%%%%%%%%%
%%%%%%%%%%%%%%%%%%%%%%%%%%%%%%%%%%%%%%%%%%%%%%%%%%%%%%%%%%%%%%%%%%%%%%%%%%%%%
\subsection{Schwarzian KP variables} 

From the previous relations for $\bu_\aR$ we can now proceed as follows
\begin{eqnarray*}
&& \bs^T\,(\chi_{\bb,\bK^\prime})^{-1}\bC\left(\chi_{\aR,\dd}\bu_\aR-\wt{V}_\aR \bu_0 \right)=
\wt{\bs}^T\,\frac{\Psi_\xi(\bK^\prime)}{\Psi_{\wt{\xi}}(\bK^\prime)}\,\frac{\Phi_{\xi+\bb}(\bK^\prime)}{\Phi_\xi(\bK^\prime)\,
\Phi_\bb(\bK^\prime)}\,e^{-\eta_\dd\bK^\prime}p_{\bK^\prime}\wt{\bu}_\aR \\ 
&& =\wt{\bs}^T\,\frac{\Phi_{\xi+\bb}(\bK^\prime)\Phi_\dd(\bK^\prime)}{\Phi_{\wt{\xi}}(\bK^\prime)\,\Phi_\bb(\bK^\prime)}\,\wt{\bu}_\aR = 
\wt{\bs}^T\,\frac{\zeta(\bK^\prime)+\zeta(\dd)+\zeta(\xi+\bb)-\zeta(\wt{\xi}+\bK^\prime+\bb)}
{\zeta(\bK^\prime)+\zeta(\bb)+\zeta(\wt{\xi})-\zeta(\wt{\xi}+\bK^\prime+\bb)}\,\wt{\bu}_\aR \nn \\ 
%\end{eqnarray*}
%\begin{eqnarray*}
&& = \wt{\bs}^T\,\left(1+\frac{\zeta(\dd)-\zeta(\bb)+\zeta(\xi+\bb)-\zeta(\xi+\dd)}
{\zeta(\bK^\prime)+\zeta(\bb)+\zeta(\wt{\xi})-\zeta(\wt{\xi}+\bK^\prime+\bb)}\right)\,\wt{\bu}_\aR=(1-\wt{V}_\aR)-
\wt{\chi}_{\bb,-\dd}\wt{S}_{\bb,\aR} \\
&& = \chi_{\aR,\dd}S_{\bb,\aR}-(1-W_\bb)\wt{V}_\aR\  , 
\end{eqnarray*}
from which we get the following relation:
\be
\label{eq:pVW} W_\bb\wt{V}_\aR=1-\wt{\chi}_{\bb,-\dd}\wt{S}_{\bb,\aR}-\chi_{\aR,\dd}S_{\bb,\aR}\   , 
\ee  
where
\begin{eqnarray}
S_{\bb,\aR}&=&\bs^T\left[ \chi_{\bb,\bK^\prime}\right]^{-1}\bC\left[ 1+\bM\bC\right]^{-1}\left[ \chi_{\aR,\bK}\right]^{-1}\boldsymbol{r}\;,\\\label{eq:S}
&=&\tbu_{\bb}\bC\left[ \chi_{\aR,\bK}\right]^{-1}\boldsymbol{r}=\bs^{T}\left[ \chi_{\bb,\bK^\prime}\right]^{-1}\bC\bu_\aR\;.\label{eq:S2}
\end{eqnarray}
Similarly we have
\bse
\be\label{eq:pVW1}
 W_\bb\wh{V}_\aR=1-\wh{\chi}_{\bb,-\ven}\wh{S}_{\bb,\aR}-\chi_{\aR,\ven}S_{\bb,\aR}\   , 
\ee
\be
 W_\bb\ol{V}_\aR=1-\ol{\chi}_{\bb,-\lambda}\ol{S}_{\bb,\aR}-\chi_{\aR,\lambda}S_{\bb,\aR}\   , 
\ee
\ese  
Note that if we multiply $\bC\left[ \wt{\chi}_{\aR,\bK}\right]^{-1}\wt\brr$ from the right hand side of Eq. \eqref{eq:tu} the
computation leads to Eq. \eqref{eq:pVW}.

Multiplying $\Phi_{\aR}(\wt\xi)\Phi_{\bb}(\xi)$ to the left hand side of \eqref{eq:pVW}, we have
\begin{eqnarray}\label{Ss}
 ( \Phi_{\aR}(\wt\xi)\wt{V}_\aR)\left(\Phi_{\bb}(\xi)W_\bb \right)&=&
-\Phi_{\bb}(\xi)\Phi_\aR(\xi)\Phi_\aR(\delta)S_{\bb,\aR}
-\Phi_{\bb}(\wt\xi)\Phi_\aR(\wt\xi)\Phi_\bb(-\delta)\wt S_{\bb,\aR}\nn\\
&&+\Phi_{\aR+\bb}(\wt\xi)\Phi_\bb(-\delta)+\Phi_{\aR+\bb}(\xi)\Phi_\aR(\delta)\nn\\
&=&\Phi_{\aR+\bb}(\wt\xi)(-p_{-\bb}+p_{-\bb}\wt\chi_{\aR,\bb}\wt S_{\bb,\aR})\nn\\
&&+\Phi_{\aR+\bb}(\xi)(p_{\aR}-p_{\aR}\chi_{\aR,\bb}S_{\bb,\aR})\;.
%&=&p_{\aR}\bZ_{\bb,\aR}-p_{-\bb}\wt\bZ_{\bb,\aR}
\end{eqnarray}
Introducing the new variable
\begin{subequations}\label{eq:w1}
\begin{eqnarray}
Z_{\bb,\aR}&=&\Phi_{\bb,\aR}(\xi)(1-\chi_{\aR,\bb}S_{\bb,\aR})\;,
\end{eqnarray}
\end{subequations}
we can write \eqref{Ss} in the form
\begin{eqnarray}\label{Ss2}
 \wt{\mathcal{V}}_\aR\mathcal{W}_\bb
=p_{\aR}Z_{\bb,\aR}-p_{-\bb}\wt Z_{\bb,\aR}\;.
\end{eqnarray}
Similarly we have
\begin{eqnarray}
 \wh{\mathcal{V}}_\aR\mathcal{W}_\bb
&=&q_{\aR}Z_{\bb,\aR}-q_{-\bb}\wh Z_{\bb,\aR}\;,\\
 \ol{\mathcal{V}}_\aR\mathcal{W}_\bb
&=&l_{\aR}Z_{\bb,\aR}-l_{-\bb}\ol Z_{\bb,\aR}\;.\label{Ss3}
\end{eqnarray}
Using the identity
\begin{eqnarray}\label{iden}
\frac{(\wt{\ol{\mathcal{V}}}_\aR\ol{\mathcal{W}}_\bb )}{(\wh{\ol{\mathcal{V}}}_\aR\ol{\mathcal{W}}_\bb)}
=\frac{(\wt{\ol{\mathcal{V}}}_\aR\wt{\mathcal{W}}_\bb )}{(\wh{\ol{\mathcal{V}}}_\aR\wh{\mathcal{W}}_\bb)}
\frac{(\wh{\wt{\mathcal{V}}}_\aR\wh{\mathcal{W}}_\bb )}{(\wh{\wt{\mathcal{V}}}_\aR\wt{\mathcal{W}}_\bb)}\;,
\end{eqnarray}
we can derive the equation
\begin{eqnarray}\label{SKP1}
\frac{(p_{\aR}\ol{Z}_{\bb,\aR}-p_{-\bb}\wt{\ol{Z}}_{\bb,\aR})}{(q_{\aR}\ol{Z}_{\bb,\aR}-q_{-\bb}\wh{\ol{Z}}_{\bb,\aR})}
=\frac{(l_{\aR}\wt{Z}_{\bb,\aR}-l_{-\bb}\wt{\ol{Z}}_{\bb,\aR})}{(l_{\aR}\wh{Z}_{\bb,\aR}-l_{-\bb}\wh{\ol{Z}}_{\bb,\aR})}\;
\;\;\frac{(p_{\aR}\wh{Z}_{\bb,\aR}-p_{-\bb}\wh{\wt{Z}}_{\bb,\aR})}{(q_{\aR}\wt{Z}_{\bb,\aR}-q_{-\bb}\wh{\wt{Z}}_{\bb,\aR})}\;,
\end{eqnarray}
which is the ``\emph{Schwarzian lattice KP equation}", first given the explicit form in \cite{DN}.
%%%%%%%%%%%%%%%%%%%%%%%%%%%%%%%%%%%%%%%%%%%%%%%%%%%%%%%%%%%%%%%%%%%%%%%
%%%%%%%%%%%%%%%%%%%%%%%%%%%%%%%%%%%%%%%%%%%%%%%%%%%%%%%%%%%%%%%%%%%%%%%%%
\\
\\
\textbf{Remark}: Let $\partial$ denote the derivative w.r.t any independent variable on 
which solely the $\rho_i$ and all the $\nu_j$ depend, but not any of the other ingredients in 
the elliptic soliton solutions. Hence $\partial$ only acts on the $\rho_i$ and $\nu_j$. 
Assuming that the latter can be written in the following form: 
\[ 
\boldsymbol r=\boldsymbol {\mathsf R}\boldsymbol e\;,\;\boldsymbol{s^T}=\boldsymbol{e^T}\boldsymbol{\mathsf S}
\;\;\;\mbox{where}\;\;\;
\boldsymbol{e^T}=(1,1,...,1).
\]
we can use the following relation for the Cauchy matrix
\[
\partial{\boldsymbol M}=(\partial\boldsymbol {\mathsf R})\boldsymbol{\mathsf R}^{-1}\boldsymbol M+
\boldsymbol M\boldsymbol{\mathsf S}^{-1}(\partial\boldsymbol{
\mathsf S}).
\]
to derive the following expression for the action of $\partial$ on the main variable $S_{\alpha,\beta}$ 
\begin{equation}\label{eq:sq_eigen} 
 \partial{S_{\alpha,\beta}}=\tbu_\beta\left[ \boldsymbol{\mathsf S}^{-1}\partial{\boldsymbol{\mathsf S}}\boldsymbol C
+\boldsymbol C\partial{\boldsymbol{\mathsf R}}\boldsymbol{\mathsf R}^{-1}\right]\bu_\alpha.
\end{equation}
This expression, in fact, is the analogue of the square eigenfunction expansion (in the sense of the 
seminal paper by Deift, Lund and Trubowitz of 1980, cf. \cite{DLT}) of the elliptic 
soliton solution of a continuous KP equation, e.g. by choosing $\partial$ to represent the partial 
derivative w.r.t. one of the continuous independent variables of the KP equation.  

In the discrete case one can derive a similar equation to \eqref{eq:sq_eigen}.  
Let``\;\;$\widetilde{}$\;\;" denote here the shift w.r.t any discrete variable on which solely 
the $\rho_i$ and $\nu_j$ depend. Using the ingredients of the previous case, one can derive the formula: 
\begin{equation}\label{eq:sq_eigen2}
 \widetilde{S}_{\alpha,\beta}-S_{\alpha,\beta}=\widetilde{\tbu}_{\beta}\left[ \boldsymbol C\boldsymbol{\widetilde{\mathsf R}}
\boldsymbol{\mathsf R}^{-1}-\boldsymbol{\widetilde{\mathsf S}}
\boldsymbol{\mathsf S}^{-1}\boldsymbol C\right]\bu_{\alpha}.
\end{equation}
which constitutes the discrete analogue to \eqref{eq:sq_eigen}. Note that the shift ``~$\wtilde{\phantom{a}}$~" may also 
represent a composite shift, or a combined shift w.r.t. multiple variables, the derivation of 
\eqref{eq:sq_eigen2} only uses the fact that the shift distributes over products of functions of the discrete variable by ~$\widetilde{fg}= \wtilde{f}\,\widetilde{g}$~ and that the discrete variable enters the solutions via the 
$\rho_i$ and the $\nu_j$. There is, however, an important difference between the formulae \eqref{eq:sq_eigen} and 
\eqref{eq:sq_eigen2}, namely that the left hand side of the latter involves also shifts of the 
eigenfunctions. We expect that these formulae may prove useful in the derivation of conservation laws for the 
discrete equations considered in this paper. 

%%%%%%%%%%%%%%%%%%%%%%%%%%%%%%%%%%%%%%%%%%%%%%%%%%%%%%%%%%%%%%%%%%%%%%%
%%%%%%%%%%%%%%%%%%%%%%%%%%%%%%%%%%%%%%%%%%%%%%%%%%%%%%%%%%%%%%%%%%%%%%%%%
\subsection{\bf Hirota form of the elliptic $N$-soliton solution} 

In this Section, we will derive some explicit formulae for the soliton solutions in terms of the $\tau$-function, which allow us to study their properties. Using the fact that the matrix $\boldsymbol M$ is actually a Cauchy matrix, the $\tau$-function can be explicitly computed by using the expansion 
$$
\tau=\det\left(\boldsymbol{1}+\boldsymbol{M}\boldsymbol{C}\right)= 1+ \sum_{i=1}^N \left|B_{i,i}\right| 
+\sum_{i<j}\left|\begin{array}{cc} B_{i,i} & B_{i,j} \\ B_{j,i} & B_{j,j} \end{array}\right| + \cdots + \det(\boldsymbol{B})\  ,  
%&& +\ld^{N-3}\sum_{i<j<k}\left|\begin{array}{ccc} M_{i,i} & M_{i,j} & M_{i,k}\\ 
%M_{j,i} & M_{j,j} & M_{j,k}\\ M_{k,i} & M_{k,j} & M_{k,k} \end{array}\right| + \cdots + \det(\boldsymbol{M})\  . \nn \\ 
$$ 
where $\bB=\bM\bC$.
\begin{lemma}
\label{CAuchyBinet}
\textbf{Cauchy-Binet formula}:
For an arbitrary $N\times M$ matrix $\boldsymbol A$ and $M \times N$ matrix $\boldsymbol B$ we have the following formula for the $N \times N$ determinant of the product \cite{Gragg}:
\begin{eqnarray}
\det_{N\times N}\left( \boldsymbol A\boldsymbol B\right)=
  \left\{ \begin{array}{rcl}
0 & \mbox{if} & M<N \\
\det(\boldsymbol A)\det(\boldsymbol B) & \mbox{if} & M=N \\
\sum_{1\leq l_1\leq ...\leq l_N\leq M}
 \det\left( \boldsymbol A_{(1,..,N)(l_1,...,l_N)}\right)
\det\left( \boldsymbol B_{(l_1,...,l_N)(1,..,N)}\right)& \mbox{if} & M>N
\end{array}\right.\nn
\end{eqnarray}
in which $\boldsymbol A_{(1,..,N)(l_1,...,l_N)}$ denotes the matrix obtained by selecting the $l_1,...,l_N$ columns from the matrix $\boldsymbol A$ and $\boldsymbol B_{(l_1,...,l_N)(1,..,N)}$ is the matrix obtained by selecting the $l_1,...,l_N$ rows from the matrix $\boldsymbol B$. $\diamond $
\end{lemma}
Using Cauchy-Binet formula, we may express the $\tau-$function in the form
\begin{eqnarray}
\tau&=&1+\sum_{i=1}^N\left( \sum_{l}^{N^\prime}M_{il}C_{li}\right)+\sum_{i<j}^N\left( \sum_{l_1<l_2}^{N^\prime}
\left| \begin{array}{cc}
 M_{il_1} & M_{il_2}  \\
M_{jl_1} & M_{jl_2} \ \end{array} \right|
\left| \begin{array}{cc}
 C_{l_1i} & C_{l_1j}  \\
C_{l_2i} & C_{l_2j} \ \end{array} \right|\right)\nn\\
&+&\sum_{i<j<s}^N\left( \sum_{l_1<l_2<l_3}^{N^\prime}
\left| \begin{array}{ccc}
 M_{il_1} & M_{il_2}& M_{il_3}  \\
M_{jl_1} & M_{jl_2}& M_{jl_3}\\
M_{sl_1} & M_{sl_2}& M_{sl_3} \end{array} \right|
\left| \begin{array}{ccc}
 C_{l_1i} & C_{l_1j}& C_{l_1s}  \\
C_{l_2i} & C_{l_2j}& C_{l_2s}\\
C_{l_3i} & C_{l_3j}& C_{l_3s} \end{array} \right|
\right)\nn\\
&+&....+\sum_{l_1<l_2<...<l_{N^{\prime}}}\det\boldsymbol{M}_{(1,2,...,N)(l_1,l_2,...,l_N)}\det\boldsymbol{C}_{(l_1,l_2,...,l_N)(1,2,...,N)}.
\end{eqnarray}
where $C_{ij}$ are the entries of the matrix $\boldsymbol C$ and $\boldsymbol{M}_{(1,2,...,N)(l_1,l_2,...,l_N)}$ denotes the matrix obtained by selecting
the $(l_1,l_2,...,l_N)$ columns from the matrix $\boldsymbol M$ and $\boldsymbol{C}_{(l_1,l_2,...,l_N)(1,2,...,N)}$ is the matrix obtained by selecting the
$(l_1,l_2,...,l_N)$ rows from $\boldsymbol{C}$.
\\
\\
Using the Frobenius formula for the relevant elliptic Cauchy determinants. Thus, from 
\begin{eqnarray}\label{eq:Frob}
\det\left(\rho_i\Phi_{\kp_i+\kp_j^\prime}(\xi)\nu_j\right)&=&\left(\prod_i\rho_i\nu_i\right)
\frac{\sg(\xi+\sum_i(\kp_i+\kp_i^\prime))}{\sg(\xi)}\,\nn\\
&&\times\frac{\prod_{i<j}\sg(\kp_i+\kp_j)\sg(\kp_i^\prime+\kp_j^\prime)}{\prod_{i,j}\sg(\kp_i-\kp_j^\prime)}\  .  
\end{eqnarray}
Introducing the notations 
\[ e^{A_{i,j}}\equiv \sg(\kp_i-\kp_j)\quad,\quad e^{\theta_i}= \rho_ie^{-\zeta(\xi)\kp_i}\  , \] 
\[ e^{E_{l_i,l_j}}\equiv \sg(\kp_{l_i}^\prime-\kp_{l_j}^\prime)\quad,\quad e^{\eta_l}= \nu_le^{-\zeta(\xi)\kp_l^\prime}\  , \] 
the Hirota formula for the $\tau$-function thus takes the form: 
\begin{eqnarray} 
\tau  &=& 1+\sum_{i=1}^N e^{\theta_i}\sum_{l=1}^{N^\prime}e^{\eta_l}\left| C^{(1 \times 1)}_{li}\right|
\frac{\sg(\xi+\kp_i+\kp_l^\prime)}{\sg(\xi)\sg(\kp_i+\kp_l^\prime)}
\nn\\
&+&\sum_{i<j}^Ne^{\theta_i+\theta_j+A_{i,j}}\,\sum_{l_1<l_2}^{N^\prime} e^{\eta_{l_1}+\eta_{l_2}+E_{l_1,l_2}}\left| C_{l_1l_2ij}^{(2\times 2)}\right|
\frac{\sg(\xi+\kp_i+\kp_j+\kp_{l_1}^\prime+\kp_{l_2}^\prime)}{\sg(\xi)
\prod_{f=1}^2\sg(\kp_i+\kp_{l_f})\sg(\kp_j+\kp_{l_f})}
\nn \\ 
&+&\sum_{i<j<s}^Ne^{\theta_i+\theta_j+\theta_s+A_{i,j}+A_{i,s}+A_{j,s}}\sum_{l_1<l_2<l_3}^{N^\prime} 
e^{\eta_{l_1}+\eta_{l_2}+\eta_{l_3}+E_{l_1,l_2}+E_{l_1,l_3}+E_{l_2,l_3}}\left| C_{l_1l_2l_3ijs}^{(3\times 3)}\right|\nn\\
&&
\times\frac{\sg(\xi+\kp_i+\kp_j+\kp_s+\kp_{l_1}^\prime+\kp_{l_2}^\prime+\kp_{l_3}^\prime)}{\sg(\xi)
\prod_{f=1}^3\sg(\kp_i+\kp_{l_f})\sg(\kp_j+\kp_{l_f})\sg(\kp_s+\kp_{l_f})}+....., \label{eq:Hirform} 
\end{eqnarray}
where $\left| C_{l_1l_2...l_Nijs...}^{(N\times N)}\right|$ is the determinant of the $N\times N$ matrix of the selected entries of $\boldsymbol C$.

The dependence of $\tau$ in \eqref{eq:Hirform} on the discrete dynamical variables $n,m,h$ comes only through the $\theta_i$ and $\eta_j$, which in turn depend on $n,m,h$ via the plane wave factors \eqref{eq:rho1} and \eqref{eq:rho22}.
%%%%%%%%%%%%%%%%%%%%%%%%%%%%%%%%%%%%%%%%%%%%%%%%%%%%%%%%%%%%%%%%%%%%%%%%%%%%%%
\section{Reduction to the lattice KdV equations}
Dimensional reductions of the lattice KP systems can be obtained by imposing certain symmetry conditions. This process could be done by imposing the condition that $T_{-\delta}\circ T_\delta u=u$ for all lattice directions. This implies that the interchange $\delta\rightarrow -\delta$ leads to a reversal of the lattice shift. 
\\
\\
We now consider the plane-wave factors shifted in the ``$\;\;\wt{}\;\;$" direction
\begin{eqnarray}
\wt\rho&=&\Phi_\delta(-\kappa )\rho\;,\\
\wt\nu &=&\Phi^{-1}_\delta(\kappa^\prime )\nu \;.
\end{eqnarray} 
We find that
\begin{equation}
T_{-\delta}\circ T_\delta (\rho \nu )=\frac{\wp (\kappa)-\wp(\delta)}{\wp(\kappa^\prime)-\wp(\delta)}\rho \nu \;.
\end{equation} 
If we take $\kappa=\kappa^\prime$ we have $T_{-\delta}\circ T_\delta (\rho \nu )=\rho \nu $.                                                              
\\
\\                      
Let's consider the lattice KP \eqref{LKP4},
\begin{equation}\label{LKP2}
\frac{\wh{\ol{w}}-\wt{\ol{w}}}{\wh{w}-\wt{w}}=\frac{\wh{\ol{w}}-\wh{\wt{w}}}{\ol{w}-\wt{w}}=\frac{\wt{\ol{w}}-\wt{\wh{w}}}{\ol{w}-\wh{w}}\;,
\end{equation}
Setting $\lambda =-\delta $, we have $T_{\lambda }\circ T_{\delta }w=\wt{\ol{w}}=w\Rightarrow \ol{w}={\hypotilde 0 w}$, where ${\hypotilde 0 w}=w(n-1,m)$, yielding
\begin{equation}\label{LKdV3}
(\wh{w}-\wt{w})({w}-\wh{\wt{w}})=\left[ (\wh{w}-\wt{w})(w-\wh{\wt{w}})\right]_{\wt{}}\;.
\end{equation}
Similarly, setting $\lambda =-\varepsilon  $, we have $T_{\lambda }\circ T_{\epsilon }w=\wh{\ol{w}}=w\Rightarrow \ol{w}={\hypohat 0 w}$, where ${\hypohat 0 w}=w(n,m-1)$, yielding
\begin{equation}\label{LKdV4}
(\wh{w}-\wt{w})({w}-\wh{\wt{w}})=\left[ (\wh{w}-\wt{w})(w-\wh{\wt{w}})\right]_{\wh{}}\;.
\end{equation}
This implies that the product on the right-hand sides of these relations is conserved in all directions, and then constant:
\begin{equation}\label{LKdV}
(\wh{w}-\wt{w})(w-\wh{\wt{w}})=\mbox{constant}\;,
\end{equation}
which is actually the elliptic version of the ``\emph{lattice KdV}" equation.
\\
\\
From \eqref{WVw3}, if we set $\lambda =-\delta $ we find that
\begin{equation}
(\wh{\wt{w}}-{w})=\frac{1}{\wh{\mathcal V}_{\aR}}\left(-p_\aR\wh{\wt{\mathcal V}}_{\aR}
-q_\aR{\mathcal V}_\aR\right)\;.\label{WVwxx}
\end{equation}
Combining this equation with \eqref{WVw4} and using the lattice KdV, we obtain 

\begin{equation}\label{MkdV}
p_{\aR}(\mathcal V_{\aR}\wh{\mathcal V}_{\aR}-\wt{\mathcal V}_{\aR}\wh{\wt{\mathcal V}}_{\aR})=q_{\aR}(\mathcal V_\aR\wt{\mathcal V}_\aR-\wh{\mathcal V}_\aR\wh{\wt{\mathcal V}}_\aR)\;,
\end{equation}
which is the ``\emph{lattice modified KdV}" equation.
\\
\\
From \eqref{BKP}, if we set $\lambda =-\delta $, we get
\begin{equation}\label{Hirotakdv1}
q_\delta {\hypotilde 0 \tau}\wh{\wt{\tau}}+p_\delta\wt{\tau}\wh{\hypotilde 0 \tau}=p_\delta \wh{\tau}\tau\;.
\end{equation}
and if we set $\lambda =-\varepsilon  $, we get
\begin{equation}\label{Hirotakdv2}
p_\delta {\hypohat 0 \tau}\wh{\wt{\tau}}+q_\delta\wh{\tau}\wt{\hypohat 0 \tau}=q_\delta \wt{\tau}\tau\;.
\end{equation}
\eqref{Hirotakdv1} and \eqref{Hirotakdv2} are the ``\emph{lattice Hirota}" equations.
\\
\\
From \eqref{Ss3}, if we set $\lambda =-\delta $ we get
\begin{equation}\label{VWpp}
\mathcal V_\aR\wt{\mathcal W}_\beta =-p_\aR\wt{Z}_{\beta ,\alpha }+p_{-\beta }{Z}_{\beta ,\alpha }\;,
\end{equation}
and if we set $\lambda =-\varepsilon  $ we get
\begin{equation}\label{VWqq}
\mathcal V_\aR\wh{\mathcal W}_\beta =-q_\aR\wh{Z}_{\beta ,\alpha }+q_{-\beta }{ Z}_{\beta ,\alpha }\;.
\end{equation}
Using the identity:
\begin{equation}\label{idenkdv}
\frac{\wt{\mathcal V}_\aR\mathcal W_\beta }{\wh{\mathcal V}_\aR\mathcal W_\beta }=\frac{\wt{\mathcal V_\aR\wh{\mathcal W}_\beta }}{\wh{\mathcal V_\aR\wt{\mathcal W}_\beta}}\;,
\end{equation}
we obtain the equation for $Z_{\beta ,\alpha }$
\begin{equation}\label{SKdV}
\frac{p_\aR Z_{\beta ,\alpha }-p_{-\beta }\wt{ Z}_{\beta ,\alpha }}{q_{\aR} Z_{\beta ,\alpha }-q_{-\beta }\wh{ Z}_{\beta ,\alpha }}=\frac{q_{-\beta }\wh{ Z}_{\beta ,\alpha }-q_\aR\wh{\wt{ Z}}_{\beta ,\alpha }}
{p_{-\beta }\wt{ Z}_{\beta ,\alpha }-p_\aR\wh{\wt{ Z}}_{\beta ,\alpha }}
\end{equation}
which is the ``\emph{lattice Schwarzian KdV}" equation. 

The dimensional reductions described here to lattice equations of the KdV class are obtained of the basis of the explicit form for the solutions described in the earlier 
sections. We would like to mention at this point that in the recent paper \cite{JamesNalini}, it was shown how to embed all equations in the ABS list 
in the Schwarzian KP equation \eqref{dSKP1}, through the canonical definition of a ``Schwarzian variable'' associated with each member of the list. 

%%%%%%%%%%%%%%%%%%%%%%%%%%%%%%%%%%%%%%%%%%%%%%%%%%%%%%%%%%%%%%%%%%%%%%%%%%%%%%%%%%%%%%%%%%%%%%%%%%%%%
\section{The continuum limit}
In previous Sections, we derive the fully discrete KP equations. The derivation of intermediate and full continuum limits of the lattice KP systems were studied in \cite{NCWQ,F5}, as well as \cite{CWH,WC2,WC} where the derivation of hierarchy arising from the lattice KP equations was systematically studied. We will consider continuum limit on the level of the elliptic soliton solution. We will give the detail only for the lattice KP \eqref{LKP4} and present the final results for the other lattice KP equations.
\\
\\
We find out that it is more convenient to express the lattice KP equation \eqref{LKP4} in the following form
\begin{eqnarray}\label{KPm}
&&\frac{\left(\wt{\ol{w}}_0-\wh{\ol{w}}_0+\zeta(\xi +\epsilon +\lambda )-\zeta(\xi+\delta+\lambda)-\zeta(\epsilon)+\zeta(\delta) \right)}
{\left(\wt{{w}}_0-\wh{{w}}_0+\zeta(\xi +\epsilon )-\zeta(\xi+\delta)-\zeta(\epsilon)+\zeta(\delta) \right)}\nn\\
&&\;\;\;\;\;=\frac{\left(\wh{\wt{w}}_0-\wh{\ol{w}}_0+\zeta(\xi +\epsilon +\lambda )-\zeta(\xi+\delta+\lambda)-\zeta(\lambda )+\zeta(\delta) \right)}
{\left(\wt{{w}}_0-\ol{{w}}_0+\zeta(\xi +\lambda  )-\zeta(\xi+\delta)-\zeta(\lambda )+\zeta(\delta) \right)}\;,
\end{eqnarray}
where $w_0=\boldsymbol{s}^T\boldsymbol{C}(1+\boldsymbol{M}\boldsymbol{C})^{-1}\boldsymbol{r}$.
\\
\\
\textbf{First continuum limit}:\textbf{ The skew limit}.
We start to perform the continuum limit by making change of discrete variables $(n,m)\rightarrow (N=n+m,m)$ and then taking the limit 
\[
\epsilon -\delta =\eta\rightarrow 0\;,\;m\rightarrow \infty \;,\;n\rightarrow -\infty \;,\;\;\mbox{s.t}\;\; m\eta\rightarrow \tau \;\;\mbox{and}\;\;N\;\mbox{fixed}\;.
\]
The plane wave functions \eqref{eq:rho1} and \eqref{eq:rho22} become
\bea
 \rho(\kp)&=&\left(e^{-\zeta(\dd)\kp}\,p_{-\kp}\right)^{~N~} 
\left(e^{-\zeta(\lambda)\kp}\,l_{-\kp}\right)^{~h~}e^{\zeta(\xi)\kp+\zeta(\delta -\kp)\tau }\rho_{0,0,0}(\kp), \label{rho1}\\
 \nu(\kp^\prime)&=&\left(e^{\zeta(\dd)\kp^\prime}\,p_{\kp^\prime}\right)^{~-N~} 
\left(e^{\zeta(\lambda)\kp^\prime}\,l_{\kp^\prime}\right)^{~-h~}e^{\zeta(\xi)\kp^\prime-\zeta(\delta +\kp^\prime)\tau}\rho_{0,0,0}(\kp^\prime),\label{rho22}
\eea 
and $w_0[n,m,h]\rightarrow w_0[N,\tau ,h]$. Then $w_0[N,\tau ,h]$ satisfies the differential-difference equation
\bea
\frac{ \frac{\partial\wt{\ol{w}}_0}{\partial\tau }+\wp (\xi+\delta +\lambda )-\wp (\delta)}
{ \frac{\partial\wt{w}_0}{\partial \tau }+\wp (\xi+\delta )-\wp (\delta)  }
=\frac{
\wt{\wt{w}}_0-\wt{\ol{w}}_0+\zeta(\xi+\delta+\lambda)-\zeta(\xi+2\delta)-\zeta(\lambda)+\zeta(\delta)}
{\wt{w}_0-\ol{w}_0+\zeta(\xi+\lambda)-\zeta(\xi+\delta)-\zeta(\lambda)+\zeta(\delta)}.
\eea 
\\
\\
\textbf{The second limit}.
Next, we perform the limit on the discrete variable $h$:
\[
\lambda\rightarrow 0\;,\;\;h\rightarrow \infty \;,\;\;\;\mbox{s.t.}\;\;\gamma =\lambda h\;\;\mbox{fixed}.
\]
The plane wave functions \eqref{rho1} and \eqref{rho22} become
\bea
 \rho(\kp)&=&\left(e^{-\zeta(\dd)\kp}\,p_{-\kp}\right)^{~N~} 
e^{\zeta(\xi)\kp+\zeta(\delta -\kp)\tau-\zeta(\kp)\gamma  }\rho_{0,0,0}(\kp), \label{rho1a}\\
 \nu(\kp^\prime)&=&\left(e^{\zeta(\dd)\kp^\prime}\,p_{\kp^\prime}\right)^{~-N~} 
e^{\zeta(\xi)\kp^\prime-\zeta(\delta +\kp^\prime)\tau+\zeta(\kp^\prime)\gamma}\rho_{0,0,0}(\kp^\prime),\label{rho22a}
\eea 
and $w_0[N,\tau ,h]\rightarrow w_0[N,\tau ,\gamma ]$. Then $w_0[N,\tau ,\gamma ]$ now satisfies the partial differential-difference equation
\begin{equation}
-\left(\frac{\partial w_{0 }}{\partial\gamma \partial\tau } +\wp^\prime(\xi)\right)
=\left( \frac{\partial w_0}{\partial \tau} +\wp(\xi)-\wp(\delta)\right)\left( \wt{w}_0-2w_0
+{\hypotilde 0 w}_0+2\zeta(\xi)-\zeta(\xi+\delta)-\zeta(\xi-\delta)\right)\;.
\end{equation}
\\
\\
\textbf{The third limit: The Full limit}.
This limit can be obtained by investigating the limiting behavior of the plane wave function. From \eqref{rho1a}, we have
\begin{eqnarray}
&&\left(e^{-\zeta(\dd)\kp}\,p_{-\kp}\right)^{~N~} 
e^{\zeta(\xi)\kp+\zeta(\delta -\kp)\tau-\zeta(\kp)\gamma  }\nn\\
&&\rightarrow \exp\left(N\ln\left|\sigma(\kp-\delta)\right|-\ln\left|\sigma(\kp)\right|+\zeta(\delta -\kp)\tau-\zeta(\kp)\gamma+\zeta(\xi)\kp\right) \nn\\
&&\rightarrow \exp\left( N\left(-\delta \zeta(\kp)-\frac{\delta ^2}{2}\wp(\kp)+\frac{\delta ^3}{6}\wp^\prime(\kp)+... \right)-\zeta(\kp)\gamma \right.\nn\\
&&\;\;\;\;\;\;\;\;\;\;\;\left.+\tau \left(-\zeta(\kp)-\delta\wp(\kp)-\frac{\delta^2}{2}\wp^\prime(\kp)+....\right)+\zeta(\xi)\kp\right)\nn\\
&&\rightarrow \exp\left(\zeta(\xi)\kp+\zeta(\kp)x+\wp(\kp)y+\wp(\kp)^\prime(\kp)t+.....\right)\;,
\end{eqnarray}
where we define $b=N\delta$ and
\begin{equation}
x=-b-\tau-\gamma\;,\;\;\;y=-\frac{\delta b}{2}-\tau\delta\;,\;\;\;\mbox{and}\;\;\;t=\frac{\delta^2b}{6}-\frac{\delta^2\tau}{2}\;.
\end{equation}
Applying the Taylor expansions, we have
\begin{equation}\label{wt}
\wt{w}_0=w_0+\delta \frac{\partial w_0}{\partial b}+\frac{\delta^2}{2}\frac{\partial^2 w_0}{\partial b^2}+\frac{\delta^3}{6}\frac{\partial^3 w_0}{\partial b^3}+\frac{\delta^4}{24}\frac{\partial^4 w_0}{\partial b^4}+....\;\;,
\end{equation}
where $b=b_0+N\delta$ and using the chain rule formulae, we have
\begin{subequations}\label{wcon}
\begin{eqnarray}
\frac{\partial w_0}{\partial \gamma } &=&-\frac{\partial w_0}{\partial x}\;,\\
\frac{\partial w_0}{\partial \tau }&=&-\frac{\partial w_0}{\partial x}-\delta\frac{\partial w_0}{\partial y}-\frac{\delta^2}{2}\frac{\partial w_0}{\partial t}\;,\\
\frac{\partial^2 w_0}{\partial \gamma \partial \tau }&=&\frac{\partial^2 w_0}{\partial x^2}+\delta\frac{\partial^2 w_0}{\partial x\partial y}+\frac{\delta^2}{2}\frac{\partial^2 w_0}{\partial x\partial t}\;.
\end{eqnarray}
\end{subequations}
Using \eqref{wt} and \eqref{wcon}, we recover the ``\emph{potential KP}" equation in order $\mathcal{O}(\delta^2)$:
\begin{equation}
[w_0]_{xt}=6[w_0]_{x}[w_0]_{xx}+\frac{3}{2}[w_0]_{yy}
+\frac{1}{2}[w_0]_{xxxx}\;,
\end{equation}
where $w_0[N,\tau ,\gamma ]\rightarrow w_0[x,y ,t ]$.
\\

At the end of this Section, we would like to mention that the continuum limit can be performed on the rest of KP equations namely the modified and Schwarzian KP equations. We will give a list of the results in the following
\\
\\
\emph{The discrete modified KP equation}:
\\
\begin{eqnarray}\label{mKPm}
\frac{\wh{\chi}_{\alpha ,\delta }\wh{V}_\alpha- \wt{\chi}_{\alpha ,\epsilon  }\wt{V}_\alpha}{\wh{\wt{V}}_\alpha }
+\frac{\ol{\chi}_{\alpha ,\epsilon  }\ol{V}_\alpha- \wh{\chi}_{\alpha ,\lambda   }\wh{V}_\alpha}{\wh{\ol{V}}_\alpha }
+\frac{\wt{\chi}_{\alpha \lambda  }\wt{V}_\alpha- \ol{\chi}_{\alpha ,\delta   }\ol{V}_\alpha}{\wt{\ol{V}}_\alpha }=0\;,
\end{eqnarray}
where $V_\alpha=1-\bs^T\bC(1+\bM\bC)^{-1}\left(\chi_{-\alpha,\boldsymbol{K}}\right)^{-1}\boldsymbol{r}$. Taking the skew limit, we have the differential-difference equation
\begin{eqnarray}\label{mKPm2}
\frac{\partial \wt{V}_{\alpha }}{\partial \tau}\wt{\ol{V}}_\alpha 
\left( \wt{\chi}_{\alpha ,\delta }\wt{\ol{V}}_\alpha-\wt{\chi}_{\alpha ,\lambda }\wt{\wt{V}}_\alpha  \right)
+\frac{\partial \wt{\ol{V}}_\alpha }{\partial \tau}\wt{\wt{V}}_\alpha \left( \wt{\chi}_{\alpha ,\lambda }\wt{V}_\alpha -\ol{\chi}_{\alpha ,\delta }\ol{V}_\alpha \right)=0\;,
\end{eqnarray}
where $V_\alpha [n,m,h]\rightarrow V_\alpha [N,\tau,h]$.
\\
\\
Taking another limit on the discrete variable $h$, we obtain the partial differential-difference equation
\begin{eqnarray}\label{mKPm3}
\frac{\partial^2 \wt{V}_\alpha }{\partial\gamma \partial\tau }=\frac{\frac{\partial\wt{V}_\alpha }{\partial \tau}\frac{\partial\wt{V}_\alpha }{\partial \gamma }}{\wt{V}_\alpha }
-\frac{\frac{\partial\wt{V}_\alpha }{\partial \tau}}{\wt{V}_\alpha}\left( \chi_{\alpha }\wt{V}_\alpha-\chi_{\alpha,\delta}V_\alpha \right)
+\frac{\frac{\partial\wt{V}_\alpha }{\partial \tau}}{\wt{\wt{V}}_\alpha}\left( \wt{\chi}_{\alpha,\delta}\wt{V}_\alpha-\chi_{\alpha }\wt{\wt{V}}_\alpha \right)\;,
\end{eqnarray}
where $\chi_\alpha =\zeta(\alpha )+\zeta(\xi+\delta)-\zeta(\xi+\alpha+\delta)$ and $V_\alpha [N,\tau,h]\rightarrow V_\alpha [N,\tau,\gamma ] $.
\\
\\
Taking the full limit, we obtain the modified KP equation
\begin{eqnarray}\label{mKPm3}
[V_\alpha ]_{xt}=[V_\alpha ]_{xxx}+3[V_\alpha ]_{yy}-6[V^2_\alpha ]_x[V_\alpha ]_{xx}-6[V_\alpha ]_y[V_\alpha ]_{xx}\;,
\end{eqnarray}
where $V_\alpha [N,\tau,\gamma]\rightarrow V_\alpha [x,y,t ] $.
\\
\\
\emph{The lattice Schwarzian KP equation}:
\begin{eqnarray}\label{SKPm1}
&&\frac{\left(1-\ol{\chi}_{\alpha ,\delta }\ol{{S}}_{\beta ,\alpha}- \wt{\ol{\chi}}_{\beta  ,-\delta   }\wt{\ol{{S}}}_{\beta ,\alpha}\right)
\left( 1-\wh{\chi}_{\alpha ,\lambda  }\wh{{S}}_{\beta ,\alpha}- \wh{\ol{\chi}}_{\beta  ,-\lambda    }\wh{\ol{{S}}}_{\beta ,\alpha}\right)
\left( 1-\wt{\chi}_{\alpha ,\epsilon   }\wt{{S}}_{\beta ,\alpha}- \wh{\wt{\chi}}_{\beta  ,-\epsilon     }\wh{\wt{{S}}}_{\beta ,\alpha}\right)}
{\left(1-\wt{\chi}_{\alpha ,\lambda  }\wt{{S}}_{\beta ,\alpha}- \wt{\ol{\chi}}_{\beta  ,-\lambda    }\wt{\ol{{S}}}_{\beta ,\alpha}\right)
\left( 1-\wh{\chi}_{\alpha ,\delta  }\wh{{S}}_{\beta ,\alpha}- \wh{\wt{\chi}}_{\beta  ,-\delta     }\wh{\wt{{S}}}_{\beta ,\alpha}\right)
\left( 1-\ol{\chi}_{\alpha ,\epsilon   }\ol{{S}}_{\beta ,\alpha}- \wh{\ol{\chi}}_{\beta  ,-\epsilon     }\wh{\ol{{S}}}_{\beta ,\alpha}\right)}\nn\\
&&=1\;,
\end{eqnarray}
where $S_{\bb,\aR}=\bs^T\left[ \chi_{\bb,\bK^\prime}\right]^{-1}\bC\left[ 1+\bM\bC\right]^{-1}\left[ \chi_{\aR,\bK}\right]^{-1}\boldsymbol{r}$. Taking the skew limit, we derive the differential-difference equation
\begin{eqnarray}\label{SKPm2}
&&\;\;\frac{(\wp(\wt{\ol{\xi}}+\alpha)-\wp(\wt{\xi}))\wt{S}_{\beta ,\alpha }+(\wp(\wt{{\xi}}+\beta )-\wp(\wt{\ol{\xi}}))\wt{\ol{S}}_{\beta ,\alpha }
+\wt{\chi}_{\alpha ,\lambda }
\frac{\partial{\wt{S}_{\beta ,\alpha }}}{\partial\tau }
+\wt{\ol{\chi}}_{\beta ,-\lambda }
\frac{\partial{\wt{\ol{S}}_{\beta ,\alpha }}}{\partial\tau }}
{1-\wt{\chi}_{\alpha ,\lambda }\wt{S}_{\beta ,\alpha }
-\wt{\ol{\chi}}_{\beta ,-\lambda }\wt{\ol{S}}_{\beta ,\alpha }}\nn\\
%%%%%%
&&+
\frac{(\wp(\wt{\ol{\xi}}+\alpha)-\wp(\delta ))\ol{S}_{\beta ,\alpha }-(\wp(\delta)+\wp(\wt{\ol{\xi}}))\wt{\ol{S}}_{\beta ,\alpha }
+\wt{\ol{\chi}}_{\beta ,-\delta  }
\frac{\partial{\wt{\ol{S}}_{\beta ,\alpha }}}{\partial\tau }}
{1-\ol{\chi}_{\alpha ,\delta  }\ol{S}_{\beta ,\alpha }
-\wt{\ol{\chi}}_{\beta ,-\delta  }\wt{\ol{S}}_{\beta ,\alpha }}\nn\\
%%%%%%
&&+
\frac{(\wp(\delta )-\wp(\wt{\xi}))\wt{S}_{\beta ,\alpha }+(\wp(\delta)+\wp(\wt{\xi}+\beta ))\wt{\wt{S}}_{\beta ,\alpha }
+\wt{\ol{\chi}}_{\alpha  ,\delta  }
\frac{\partial{\wt{\bS}_{\beta ,\alpha }}}{\partial\tau }}
{1-\wt{\chi}_{\alpha ,\delta  }\wt{S}_{\beta ,\alpha }
-\wt{\wt{\chi}}_{\beta ,-\delta  }\wt{\wt{S}}_{\beta ,\alpha }}=0\;,
\end{eqnarray}
where $S_{\bb,\aR}[n,m,h]\rightarrow S_{\bb,\aR}[N,\tau ,h]$.
\\
\\
Taking another limit on the discrete variable $h$, we have the partial differential-difference equation
\begin{eqnarray}\label{SKPm3}
&&\;\;\frac{[\wp(\xi+\beta )+\wp(\xi+\alpha )-2\wp(\xi)] S_{\beta ,\alpha }
+[\chi_{\alpha ,\delta  }+\chi_{\beta  ,-\delta  }]
\frac{\partial{S_{\beta ,\alpha }}}{\partial\tau }
-\frac{\partial^2{S_{\beta ,\alpha }}}{\partial\tau\partial\gamma}}
{1-[\chi_{\alpha ,\delta  }+\chi_{\beta  ,-\delta  }]S_{\beta ,\alpha }
+\frac{\partial S_{\beta ,\alpha }}{\partial\gamma}}\nn\\
%%%%%%
&&+
\frac{[\wp(\delta )-\wp(\xi)]S_{\beta ,\alpha }
+[\wp(\delta)+\wp(\xi+\beta )]\wt{S}_{\beta ,\alpha }
+\chi_{\alpha  ,\delta  }
\frac{\partial S_{\beta ,\alpha }}{\partial\tau }}
{1-\chi_{\alpha ,\delta  }S_{\beta ,\alpha }
-\wt{\chi}_{\beta ,-\delta  }\wt{S}_{\beta ,\alpha }}\nn\\
%%%%%%
&&+
\frac{[\wp(\xi +\alpha  )-\wp(\delta )]\hypotilde 0 {S}_{\beta ,\alpha }-[\wp(\delta)+\wp(\hypotilde 0 {\xi} )]S_{\beta ,\alpha }
+\chi_{\beta   ,-\delta  }
\frac{\partial S_{\beta ,\alpha }}{\partial\tau }}
{1-\hypotilde 0 {\chi}_{\alpha ,\delta  }\hypotilde 0 {S}_{\beta ,\alpha }
-\chi_{\beta ,-\delta  }S_{\beta ,\alpha }}=0\;,
\end{eqnarray}
where $S_{\bb,\aR}[N,\tau ,h]\rightarrow S_{\bb,\aR}[N,\tau ,\gamma ]$.
\\
\\
Taking the full limit, we obtain the Schwarzian KP equation
\begin{eqnarray}\label{SKPm4}
\frac{\partial}{\partial x}\left( \frac{[S_{\bb,\aR}]_t-[S_{\bb,\aR}]_{xxx}}{[S_{\bb,\aR}]_x}+\frac{2[S_{\bb,\aR}]^2_y-3[S_{\bb,\aR}]^3_{xx}}{2[S_{\bb,\aR}]^2_{x}}\right)+\frac{\partial}{\partial y}\left( \frac{[S_{\bb,\aR}]_{y}}{[S_{\bb,\aR}]_{x}}\right)=0\;,
\end{eqnarray}
where $S_{\bb,\aR}[N,\tau ,h]\rightarrow S_{\bb,\aR}[x,y,t]$.

%%%%%%%%%%%%%%%%%%%%%%%%%%%%%%%%%%%%%%%%%%%%%%%%%%%%%%%%%%%%%%%%%%%%%%%%%%%%%%%%%%%%%%%%%%%%%%%55
\section{Summary}
In this paper, we established the explicit form of a class of elliptic soliton solutions for all the lattice KP equations, based on a construction 
using elliptic Cauchy matrices denoted by $\boldsymbol M$. urthermore, the construction exhibits numerous relations between the various lattice equations, 
as well as corresponding Lax pairs.  The explicit form for the corresponding $\tau$-function depends crucially on the coefficient matrix $\boldsymbol C$, which 
opens the way to classify the various different lattice soliton behaviours according to the Schubert decompositions of the corresponding 
Grassmannians, following similar work in the continuous case by Kodama and Chalravarty, \cite{Kodama}. Several reductions were considered: {\it i)} 
dimensional reduction to KdV lattice systems, and {\it ii)} continuum limits to the semidiscrete and fully continuous KP equations. For all these equations 
the corresponding elliptic soliton solutions are derived in parallel. Furthermore, the result in \eqref{eq:Hirform} can be simplified to the cases of trigonometric/hyperbolic by taking: $\sigma(x)\mapsto \sin(x)$ or $\sigma(x)\mapsto \sinh(x)$ and for the rational case we have $\sigma(x)\mapsto x$.

There are well-established connections between soliton solution of integrable PDE and the integrable many body systems \cite{R1,R2}. This can  be made most 
explicit in he rational and trigonometric/hyperbolic cases. The elliptic case of this correspondence is more difficult to establish, but we expect 
that the elliptic solitons which we have studied here can be connected to elliptic case of the discrete-time Ruijsenaars model constructed in \cite{Frank1}, 
but at this juncture this is still conjectural. Nevertheless, in \cite{Rinthesis,RinRS}, an explicit connection between the rational discrete-time RS system 
and the KP lattice was established using the solution structure of the former model.

%%%%%%%%%%%%%%%%%%%%%%%%%%%%%%%%%%%%%%%%%%%%%%%%%%%%%%%%%%%%%%%%%%%%%%%%
%%%%%%%%%%%%%%%%%%%%%%%%%%%%%%%%%%%%%%%%%%%
\begin{acknowledgements}
SYK was supported by the Royal Thai Government and FWN is supported by a Royal Society/Leverhulme Trust senior research fellowship. 
\end{acknowledgements}
%%%%%%%%%%%%%%%%%%%%%%%%%%%%%%%%%%%%%%%%%%%%

%%%%%%%%%%%%%%%%%%%%%%%%%%%%%%%%%%%%%%%


\begin{thebibliography}{}
%AAAAA
\bibitem{Moser} 
Airault H, McKean H and Moser J,
\emph{Rational and Elliptic Solutions fo the Kortewegde Vries Equation and a Related Many-Body Problem}, 
Commun Pure Appl. Math, \textbf{30}, 95(1997).
%
\bibitem{JamesNalini}
Atkinson J and Joshi N,
\emph{The Schwarzian variable associated with discrete KdV-type equations}, 
arXiv:1010.1916v1(2010).
%
%
\bibitem{Q4}
Atkinson J and Nijhoff  F W,
\emph{A constructive approach to the soliton solutions of integrable quadrilateral lattice equations}, Commun Maths Phys, \textbf{299}, 283(2010).
%
%
\bibitem{ABS}
Adler V E, Bobenko A I, Suris Y B, 
\emph{Classification of integrable equations on quad-graphs, the consistency approach}, 
Commun. Math. Phys, {\bf 233}, 513(2003).
\bibitem{ABS2}
Adler V E, Bobenko A I, Suris Y B, 
\emph{Classification of integrable discrete equations of octahedron type}, 
Int Math Res Notices, \textbf{17}(2011).
%
%%%%%%%%%%%%%%%%%%%%%%%%%%%%%%%%%%%%%%%%%%
%%%%%%%%%%%%%%%%%%%%%%%%%%%%%%%%%%%%%%%%%%
%BBBBBBBB
%\bibitem{BS} 
%Bobenko A I, Suris Y B, 
%\emph{Discrete Differential Geometry: Integrable Structure},
%Graduate Studies in Mathematics, Vol. 98, AMS, Providence, xxiv+404 2008.
%
\bibitem{BK} 
Bogdanov L V, Konopelchenko B G, 
\emph{Analytic-bilinear approach to integrable hierarchies. II. Multicomponent KP and 2D Toda lattice hierarchies},
J. Math. Phys, \textbf{39}, 4701(1998).
%
%%%%%%%%%%%%%%%%%%%%%%%%%%%%%%%%%%%%%%%%%%%%%
\bibitem{CWH}
Capel W H, Wiersma L G and Nijhoff W F, \emph{Linearizing integral transform for the multicomponent lattice KP}, Physica, \textbf{138A}, 76(1986).
%%%%%%%%%%%%%%%%%%%%%%%%%%%%%%%%%%%%%%%%%%%
\bibitem{DLT}
Deift P, Lund F and Trubowitz E, 
\emph{Nonlinear wave equations and constrained harmonic motion},, 
Commun. Math. Phys, \textbf{74}, 141(1980).
%%%%%%%%%%%%%%%%%%%%%%%%%%%%%%%%%%%%%%%%%%%
%DDDDD
\bibitem{Doliwa1}
Doliwa A,
\emph{Desargues maps and the Hirota-Miwa equation}, 
Proc. Royal Soc. A, \textbf{466}, 1177(2010).
%
%\bibitem{Doliwa2}
%Doliwa A,
%\emph{The affine Weyl group symmetry of Desargues maps and of the non-commutative Hirota-Miwa system}, 
%Phys. Lett. A, \textbf{375}, 1219(2011).
%
\bibitem{DN}
Dorfman I Ya, Nijhoff W F, 
\emph{On a (2+1)-dimensional version of the Krichever-Novikov equation}, 
Phys. Lett, \textbf{157A}, 107(1991).
%%%%%%%%%%%%%%%%%%%%%%%%%%%%%%%%%%%%%%%%
%%%%%%%%%%%%%%%%%%%%%%%%%%%%%%%%%%%%%%%%%
%GGGGGGG
%
\bibitem{Gragg} 
Gragg W B, 
\emph{The Pade Table and Its Relation to Certain Algorithms of Numerical Analysis} 
SIAM Review, \textbf{14}, 1(1972).
%
%%%%%%%%%%%%%%%%%%%%%555
%%%%%%%%%%%%%%%%%%%%%%%%
%%HHHHHHHH
\bibitem{Hirota} 
Hirota R, 
\emph{Discrete Anologue of a Generalised Toda Equation}, 
J. Phys. Soc. Japan, \textbf{50}, 3785(1981).
%
%\bibitem{Hirotabook} 
%Hirota R, \emph{The direct method in soliton theory}, Cambridge University Press, 2008.
%
%\bibitem{Hay}
%Hay M, Kajiwara K and Masuda T, 
%\emph{Bilinearization and Special Solutions to the Discrete Schwarzian KdV Equation}, 
%J Math-for-Industry, \textbf{3}, 53(2011).
%
%%%%%%%%%%%%%%%%%%%%%%%%%%%%%%%%%%%%%%%%%%%%%%%
%%%%%%%%%%%%%%%%%%%%%%%%%%%%%%%%%%%%%%%%%%%%%%%%
%KKKKKK
\bibitem{Kacbook} 
Kac G V, \emph{Infinite dimensional Lie algebras}, Cambridge University Press, 1994.
\bibitem{KS}
King A D, Schief W K,
\emph{Tetrahedra, octahedra and cubo-octahedra: integrable geometry of multi-ratios}, 
J. Phys. A, \textbf{36}, 785(2003).
%
\bibitem{KS2}
Konopelchenko B G, Schief W K, 
\emph{Menelaus’ theorem, Clifford configurations and inversive geometry of the Schwarzian KP hierarchy}, 
J. Phys. A, \textbf{35}, 6125(2002).
%
%\bibitem{KLWZ}
%Krichever I, Lipan O, Wiegmann P, and Zabrodin Z, 
%\emph{Quantum integrable models and discrete classical Hirota equations}, 
%Comm. Math. Phys, \textbf{188}, 267(1997).
%
\bibitem{Kodama}
Kodama Y, 
\emph{Young diagrams and N-soliton solutions of the KP equation}, 
J. Phys. A, \textbf{37}, 11169(2004).
%

%%%%%%%%%%%%%%%%%%%%%%%%%%%%%%%%%%%%%
%LLLLLL
\bibitem{Lipan} 
Lipan O, Wiegmann P and Zabrodin Z, 
\emph{Fusion Rules for Quantum transfer Matrices as Dynamical Systems on Grassmann Manifolds}, 
Preprint UNiv. of Chicago(1997).
%
%\bibitem{LNQ}
%Lobb S B, Nijhoff F W, and Quispel G R W, 
%\emph{Lagrangian multiform structure for the lattice KP system}, 
%J. Phys. A, \textbf{42}, 472002(2009).
%
%%%%%%%%%%%%%%%%%%%%%%%%%%%%%%%%%%%%%%
\bibitem{Miwa} 
Miwa T, 
\emph{On Hirota's difference Equations}, 
Proc. Japan Acad, \textbf{58}, 9(1982).

%%%%%%%%%%%%%%%%%%%%%%%%%%%%%%
%NNNNNN
%\bibitem{FrankN8} 
%Nijhoff F W, Ragnisco O and Kuznetsov V B, 
%\emph{Integrable Time-Discretization of the Ruijsenaars-Schneider Model}, 
%Commun. Math. Phys, \textbf{176}, 681(1996).
\bibitem{NAH}
Nijhoff W F, Atkinson J and Hietarinta J, 
\emph{ Soliton Solutions for ABS Lattice Equations: I Cauchy Matrix Approach}, 
J. Phys. A: Special Issue dedicated to the Darboux Days, \textbf{42}, 404005(2009).
\bibitem{FrankJames} 
Nijhoff W F and Atkinson J, 
\emph{Elliptic N-soliton solutions of ABS lattice equations}, 
Int. Math. Res. Notices, \textbf{20}, 3837(2010).
%\bibitem{NC}
%Nijhoff F W, Capel H W, 
%\emph{The direct linearisation approach to hierarchies of inte-
%grable PDEs in 2 + 1 dimensions: I. Lattice equations and the differential-difference hierarchies}, 
%Inv. Problems, \textbf{6}, 567(1990).
%
\bibitem{NCW}
Nijhoff W F, Capel H W, and Wiersma G L, 
\emph{Integrable lattice systems in two and
three dimensions}, In: Geometric Aspects of the Einstein Equations and Integrable
Systems, Ed. R. Martini, Lecture Notes in Physics, Berlin/New York, Springer Verlag, 263(1985).
%
\bibitem{NCWQ}
Nijhoff W F, Capel H W, Wiersma G L, and Quispel G R W, 
\emph{Backlund transformations and three-dimensional lattice equations}, 
Phys. Lett, \textbf{105A}, 267(1984).
%
%
\bibitem{F5} Nijhoff F W, Capel H W  and Wiersma G L, \emph{Integrable Lattice Systems in Two and Three Dimensions}, Ed. R. Martini, 
in: Geometric Aspects of the Einstein Equations and Integrable Systems, Lecture Notes in Physics, pp. 263--302, Berlin/New York, Springer Verlag(1985).
%

%
%\bibitem{Franktau}
%Nijhoff F W, 
%\emph{The direct linearizing transform for the $\tau$ function in three-dimensional lattice equations}, 
%Phys. Lett. A, \textbf{110}, 10(1985).
%
%\bibitem{FrankSchBook}
%Nijhoff F W, \emph{On some ``Schwarzian" equation and their discrete analogues}. In Algebraic Aspects of Integrable Systems: In Memory of Irene Dorfman. Editors: A. S. Fokas and I. M. Gelfand. Progress in Nonlinear Differential Equations. Volume 26(1996).
%
\bibitem{Frank1} 
Nijhoff W F, Ragnisco O and Kuznetsov V B, 
\emph{Integralble Time-Discretization of the Ruijsenaars-Schneider Model},
 Commun. Math. Phys, \textbf{176}, 681(1996).
\bibitem{NY}
Noumi M and Yamada Y, \emph{Affine Weyl groups, discrete dynamical systems and Painlev\'e equations}, Commun. Math. Phys, {\bf 217} 315(2001).  
%%%%%%%%%%%%%%%%%%%%%%%
%RRRRRR
\bibitem{R1} 
Ruijsenaars S N M and Schneider H, 
\emph{A New Class of Integrable Systems and Its relation to Solitons}, 
Ann. Phys, \textbf{170}, 370(1986).
\bibitem{R2} 
Ruijsenaars S N M,
\emph{Complete integrability of Relativistic Calogero-Moser systems and Elliptic Function Identities}, 
Commun. Math. Phys, \textbf{110}, 191(1987).
%
%\bibitem{RGS}
%Ruijsenaars S N M, 
%\emph{Integrable particle systems vs solutions to the KP and 2D Toda equations},
% Ann. Phys. (NY), \textbf{256}, 226(1997).
%
%
%\bibitem{RGS}
%Ramani A, Grammaticos B and Satsuma J, 
%\emph{Integrability of multidimensional discrete systems}, 
%Phys. Lett. A, \textbf{169}, 323(1992).
%
%YYYYYYYYYYYYYYY
\bibitem{Rinthesis} 
Yoo-Kong S, 
\emph{Calogero-Moser type systems, associated KP systems, and Lagrangian structure}, 
Thesis, University of Leeds(2011).
\bibitem{RinRS} 
Yoo-Kong S and Nijhoff W F,
\emph{Discrete-time Ruijsenaars-Schneider system and Lagrangian 1-form structure}, in preparation.
\bibitem{WC2}
Wiersma L G and Capel W H, \emph{Lattice equations, hierarchies and Hamiltonian structures: The Kadomtesv-Petviashvili equation}, Phys. Lett. {\bf 124A}, 124(1987).
\bibitem{WC}
Wiersma L G and Capel W H, \emph{Lattice equations, hierarchies and Hamiltonian structures: II. KP-type of hierarchies
 on 2D lattices}, Physica, {\bf 149A}, 49(1988); ibid. \emph{III. The 2D toda and KP hierarchies}, Physica, {\bf 149A}, 75(1988).  
%%%%%%%%%%%%%%%%%%%%%%%%%%
\bibitem{Zabrodin1} 
Zabrodin Z, 
\emph{Discrete Hirota's equation in Quantum Integrable Models}, 
Theor. Math. Phys, \textbf{113}, 1347(1997).

%\bibitem{F5} 
%F.W. Nijhoff, H.W. Capel and G.L. Wiersma, Integrable Lattice Systems in Two and Three Dimensions, Ed. R. Martini, in: Geometric Aspects of the Einstein Equations 
% Integrable Systems, Lecture Notes in Physics, pp. 263--302, Berlin/New York, Springer Verlag, 1985.
%\bibitem{LWZ} 
%Lipan O, Wiegmann P and Zabrodin Z 1997 {\emph{Fusion Rules for Quantum transfer Matrices as Dynamical Systems on Grassmann Manifolds}}, Preprint UNiv. of Chicago.
%
%
%\bibitem{ABS} 
%V.E. Adler. Classification of discrete integrable equations of Hirota
%type. Talk at the Workshop “Geometric Aspects of Discrete and
%Ultra-discrete
%Integrable
%Systems”,
%30.03.-03.04.2009,
%Glasgow,
%UK,
%http://www.newton.ac.uk/programmes/DIS/seminars/033014159.html.
%

%
%\bibitem{KNS}
%A. Kuniba, T. Nakanishi, and J. Suzuki. T-systems and Y-systems in integrable sys-
%tems. arXiv: 1010.1344 [hep-th], 155 pp





\end{thebibliography}
\end{document}